# A cable finite element formulation based on exact tension field for static nonlinear analysis of cable structures

Wenxiong Li*, Qikun Huang, Suiyin Chen*

College of Water Conservancy and Civil Engineering, South China Agricultural University, Guangzhou 510642, China

*Corresponding author. E-mail: leewenxiong@scau.edu.cn (W. Li), rinchan41@scau.edu.cn (S. Chen)

**ABSTRACT**

This paper presents a numerically exact cable finite element model for static nonlinear analysis of cable structures. The model derives the exact expression of the tension field using the geometrically exact beam theory coupled with the fundamental mechanical characteristics of cables. The equations for the cable elements are formulated by addressing the equilibrium conditions at the element boundaries and ensuring compatibility within the elements. Unlike previous studies that typically provide explicit expressions for cable models, this study develops a formulation that emphasizes numerical precision and broad applicability. It achieves this by deriving linearized equations with implicit expressions incorporating integrals. The present model accurately computes internal forces and deformation states, and determines the unstrained length of the cable. Additionally, it accounts for the variability in cross-sectional stiffness along the cable's length. The paper discusses solution implementations using the complete tangent matrix and element internal iterations. The effectiveness of the proposed cable elements is demonstrated through numerical examples.

## 1 Introduction

Cable structures are utilized extensively in various engineering fields, including large-span suspension bridges, cable-stayed bridges, aerial tramways, and aerospace deployable structures [1]. The complex geometric configurations and significant geometric nonlinearity amplify the complexity of numerical simulations for these structures. These complexities present significant challenges in accurately predicting the performance of cable structures and effectively managing their construction processes. Consequently, developing efficient and precise computational methods for cable structures is a critical focus within this field. Extensive research has been conducted on various methodologies to analyze cable behavior. Two primary approaches have been extensively employed: the finite element method using interpolation functions and the analytical method, which includes explicit expressions of a catenary.

The finite element method is applied to simulate the nonlinear behaviors of cable structures. Currently, the finite element models predominantly utilized for these simulations include: cable element models based on modified physical properties, truss element models, and cable element models employing high-order interpolation functions. The cable element model based on modified physical properties typically adjusts the elastic modulus to account for sag effects, following the methodology outlined in the Ernst formula and its revisions [2, 3]. However, this model does not account for hardening effects due to large displacements. As a result, this model achieves high accuracy primarily only under conditions of high cable stress and low chord angles. The truss element model simulates the nonlinear behavior of cables using multiple truss elements [4, 5]. Increasing the number of truss elements enhances the approximation to the actual solution, yet also significantly increases computational demands, which challenges its practical application in engineering. The cable element model that uses high-order interpolation functions more accurately represents the cable's shape and accounts for the sag effect, outperforming the model based on modified physical properties. It also addresses the high computational load issue associated with the truss element model. Advanced implementations of this model include elements with 3 to 6 nodes [6-9]. In addition, some cable elements with high-order interpolation functions are developed on the basis of the geometrically exact beam theory [10-20], examples being the rotation-free geometrically exact beam element [21], umbilical cables in deep-sea remotely operated vehicle systems [22], and the rotational quaternion-based geometrically exact beam element model for simulating flexible cables in the automotive industry [23]. Despite their advantages, these models still require a significant number of elements to accurately represent the cable's form, leading to considerable computational loads.

The analytical approach effectively addresses the sag effect and typically requires just a two-node element to maintain computational accuracy. O'Brien and Francis [24] first proposed this element based on the analytical expressions of the elastic catenary, demonstrating that a single analytical element can represent each cable within a structure. In this model, the overall equilibrium of a stretched cable element is achieved using Lagrangian coordinates, with the precise cable profile derived by applying boundary conditions at the cable's endpoints. Subsequent studies have enhanced the elastic catenary element by including thermal effects and diverse loading types [25-29]. For instance, Salehi Ahmad Abad et al. [30] developed an extended three-dimensional catenary element that accommodates thermal effects and distributed lateral loads across various directions. Crusells-Girona et al. [31] applied a mixed variational method in curvilinear coordinates based on the elastic catenary expressions to model cables with material and geometric nonlinearity. Impollonia et al. [32] introduced the elastic catenary theory for extensible cables under uniformly distributed loads and three-dimensional point forces. In the context of three-dimensional cable structures, Greco et al. [33] utilized the elastic catenary solution for uniform distributed loads and additional point forces, formulating the equations based on the equilibrium at free nodes and compatibility at the terminal node of each cable. Additionally, form-finding is performed using the catenary force density method. The catenary element has also been applied in the nonlinear analysis of cable-supported bridges [34, 35] and suspension bridges [36-38]. Despite these applications, the catenary element has its limitations. These limitations mainly manifest in two aspects. Firstly, the catenary element based on analytical expression of cable shape is not suitable for predicting the nonlinear behavior of cables under the varying cross-sectional stiffness along the cable axis. Secondly, the analytical expression of cable shape is usually derived under the assumption of uniformly distributed loads, which ignores the non-uniformity of self-weight caused by changes in tension, resulting in significant errors in the analysis of cables with high self-weight. Rezaiee-Pajand et al. [25] proposed the elastic hyperbolic element for nonlinear thermo-elastic analysis to address some of these disadvantages. In addition, the parabolic approach has also been utilized [39] for the analysis and design of practical cable structures, though the error of this method increases with the sag-to-span ratio. While the analytical approach offers a solution with acceptable accuracy, its applicability remains limited to scenarios with uniform cross-sectional stiffness. High-precision cable elements designed for cables with non-uniform cross-sectional stiffness along the cable axis require further research. Moreover, the elements based on analytical functions may become unstable during the solution process if the relative positions of the nodes within an element change.

Santos and Almeida [40] introduced a complementary-dual force-based finite element formulation for the geometrically exact quasi-static analysis of cable structures employing both Hookean and Neo-Hookean materials. This formulation treats the axial force as an indeterminate field quantity and integrates the nodal equilibrium constraint using the Lagrange multiplier method, facilitating the derivation of the element stiffness matrix via variational principles. With its capability to accurately represent the tension field, this element provides high computational precision for deformation analysis and internal force calculations. However, this formulation does not address scenarios where the unstrained length of the cable is the unknown variable, including specific applications like form-finding in cable structures and construction monitoring of cable systems. Therefore, further research is essential to develop a high-precision cable element that can accurately compute deformation, internal forces, and the unstrained length of cables.

This paper introduces a numerically exact cable finite element model for static nonlinear analysis of cable structures. Starting from the differential equilibrium equations and geometric relations within the framework of planar geometrically exact beam theory, the fundamental assumptions that align with the mechanical characteristics specific to cables are introduced to derive the analytical expression for the tension field of the cable. Based on the analytical tension field, the equation systems for the proposed cable elements are developed according to the equilibrium conditions at the element boundaries and compatibility conditions within the element, which are significantly different from the existing finite element methods that pose weak form of differential equations for an elastic cable. In addition, unlike existing studies that often provide explicit expressions for cable models, this work aims to offer a cable element formulation with enhanced numerical accuracy and broader applicability. This is accomplished by directly deriving linearized equations with implicit integral expressions. The present model is suitable for cables with non-uniform cross-sectional stiffness along the cable axis and utilizes a two-node element format to improve computational efficiency. The cable element model is capable of determining internal forces, deformation states, and the unstrained length of the cable. The effectiveness of the proposed cable elements is demonstrated through numerical examples.

## 2 Formulation of the cable

### 2.1 Basic assumptions

The following assumptions are adopted for the cable:

(A1) The cable is assumed to be perfectly flexible, with no flexural stiffness.

(A2) Due to its flexibility, the cable does not undergo shear deformation. Additionally, as a tension-only element, it does not withstand compressive forces.

(A3) Material properties remain linear in the deformed state.

(A4) The self-weight per unit unstrained length is constant. For a cable with a specific unstrained length, the total self-weight of the cable remains constant, while the self-weight per unit deformed length is variable.

### 2.2 Equilibrium equations and strain expressions

With respect to an orthonormal reference system with base vectors $\mathbf{g}_1 = \{1 \quad 0\}^{\mathrm{T}}$ and $\mathbf{g}_2 = \{0 \quad 1\}^{\mathrm{T}}$, **Fig. 1** shows the reference and deformed configurations of a cable, where $L_0$ and $L$ represent the unstrained length and deformed length, respectively, $\theta$ refers to the rotation of the cross-section, and $q$ represents the self-weight along the axis of unstrained cable. The centroid lines of the cable for the deformed configuration and unstrained configuration are described by $\mathbf{r}(s)$ and $\mathbf{r}_0(s)$, respectively, $s \in [0, L_0]$ is the Lagrangian coordinate, referred to as the arc-length of the unstrained cable between the generic point and the cable origin $a$, and $q(s)$ and $\bar{q}(s)$ represent the vertically distributed loads corresponding to the reference configuration (unstrained configuration) and deformed configuration, respectively. According to assumption (A4), the vertically distributed load corresponding to unstrained configuration is constant and expressed as $q(s) = q$, where $q < 0$. Specifically, $\mathbf{r}(s)$ and $\mathbf{r}_0(s)$ contain two components corresponding to $\mathbf{g}_1$ and $\mathbf{g}_2$, and they are expressed as

$$\mathbf{r}(s) = \{r_1(s) \quad r_2(s)\}^{\mathrm{T}} \tag{1}$$

$$\mathbf{r}_0(s) = \{r_{01}(s) \quad r_{02}(s)\}^{\mathrm{T}} \tag{2}$$

In the planar geometrically exact beam theory, the generalized strains $\varepsilon_G(s)$, $\gamma_G(s)$ and $\kappa_G(s)$ measuring the extension deformation, shearing deformation and curvature in the cross-sectional coordinate system are expressed as [12]

$$\varepsilon_G(s) = r_{1,s}(s)\cos\theta(s) + r_{2,s}(s)\sin\theta(s) - 1 \tag{3}$$

$$\gamma_G(s) = r_{2,s}(s)\cos\theta(s) - r_{1,s}(s)\sin\theta(s) \tag{4}$$

$$\kappa_G(s) = \theta_{,s}(s) - \theta_{0,s}(s) \tag{5}$$

where $\theta_0(s)$ is the rotation of cross-section in the reference configuration, and $(\cdot)_{,s} = \mathrm{d}(\cdot)/\mathrm{d}s$ denotes the first-order derivative with respect to $s$. Specifically, the subscript '$G$' is used for denoting the quantities corresponding to the cross-sectional coordinate system. Meanwhile, the differential equilibrium equations of the planar geometrically exact beam theory are expressed as [12]

$$N_{g_1,s}(s) - \bar{n}(s) = 0 \tag{6}$$

$$N_{g_2,s}(s) - \bar{q}(s) = 0 \tag{7}$$

$$M_{g,s}(s) - r_{2,s}(s)N_{g_1}(s) + r_{1,s}(s)N_{g_2}(s) - \bar{m}(s) = 0 \tag{8}$$

where $N_{g_1}(s)$ and $N_{g_2}(s)$ are the internal force components in two directions $\mathbf{g}_1$ and $\mathbf{g}_2$, respectively, $M_g(s)$ is the bending moment, $\bar{n}(s)$ and $\bar{q}(s)$ represent the distributed external forces along the axis in two directions $\mathbf{g}_1$ and $\mathbf{g}_2$, respectively, and $\bar{m}(s)$ refers to the distributed external moment.

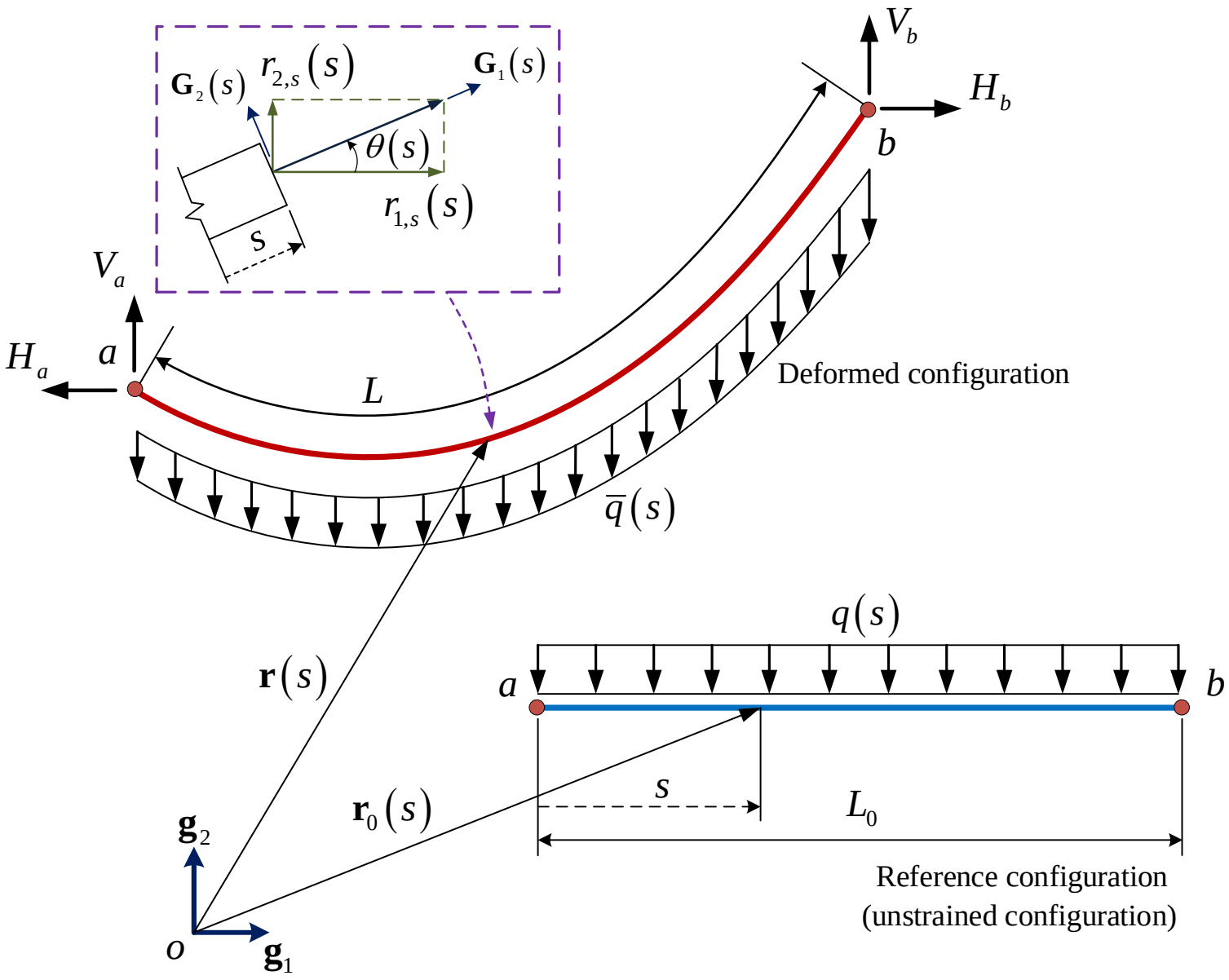

**Fig. 1**. The reference and deformed configurations of the cable.

In this study, $\bar{n}(s)$ and $\bar{m}(s)$ can be removed because only the vertically distributed load reflecting the self-weight of the cable is considered. Then, according to assumption (A1), the differential equilibrium equations of the cable can be rewritten as

$$N_{g_1,s}(s)=0 \tag{9}$$

$$N_{g_2,s}(s)-\bar{q}(s)=0 \tag{10}$$

$$r_{2,s}(s)N_{g_1}(s)=r_{1,s}(s)N_{g_2}(s) \tag{11}$$

According to assumption (A2), the following relations can be obtained

$$r_{2,s}(s)\cos\theta(s)-r_{1,s}(s)\sin\theta(s)=0 \tag{12}$$

With the relation $\cos^2\theta(s)+\sin^2\theta(s)=1$ introduced, the expressions of $\cos\theta(s)$ and $\sin\theta(s)$ can be derived as

$$\begin{aligned}\cos\theta(s)&=r_{1,s}(s)\Big/\sqrt{r_{1,s}^2(s)+r_{2,s}^2(s)}\\ \sin\theta(s)&=r_{2,s}(s)\Big/\sqrt{r_{1,s}^2(s)+r_{2,s}^2(s)}\end{aligned} \tag{13}$$

Furthermore, by substituting Eq. (13) into Eq. (3), the generalized strain $\varepsilon_G(s)$ can be expressed by $r_1(s)$ and $r_2(s)$ as

$$\varepsilon_G(s)=\sqrt{r_{1,s}^2(s)+r_{2,s}^2(s)}-1 \tag{14}$$

**2.3 Expression of the tension**

The solution of Eqs. (9) can be expressed as

$$N_{g_1}(s) = H \tag{15}$$

where $H$ represents the internal force component $N_{g_1}(s)$ at node $b$ and is regarded as one of the force parameters. It is shown that the internal force component $N_{g_1}(s)$ is constant along the cable.

The solution of Eqs. (10) can be expressed as

$$N_{g_2}(s) = V + \int_s^{L_0} \bar{q}(\xi)\,\mathrm{d}\xi = V + \int_s^{L_0} q\,\mathrm{d}\xi = V + qL_0 - qs \tag{16}$$

where $V$ represents the internal force component $N_{g_2}(s)$ at node $b$ and is regarded as another force parameter. **Fig. 2** demonstrates the relationship between $N_{g_2}(s)$ and $V$. It can be observed that $N_{g_2}(s)$ is related to the Lagrangian coordinate *s*, the unstrained length $L_0$ and the distributed load *q* under a given *V*. For a specified set of *s*, $L_0$ and *q*, the total load in the $[s, L_0]$ interval, namely $\int_s^{L_0} \bar{q}(\xi)\,\mathrm{d}\xi$, is constant and can be expressed as $\int_s^{L_0} q\,\mathrm{d}\xi$. In other words, the internal force component $N_{g_2}(s)$ in any deformed state is determined by *V*.

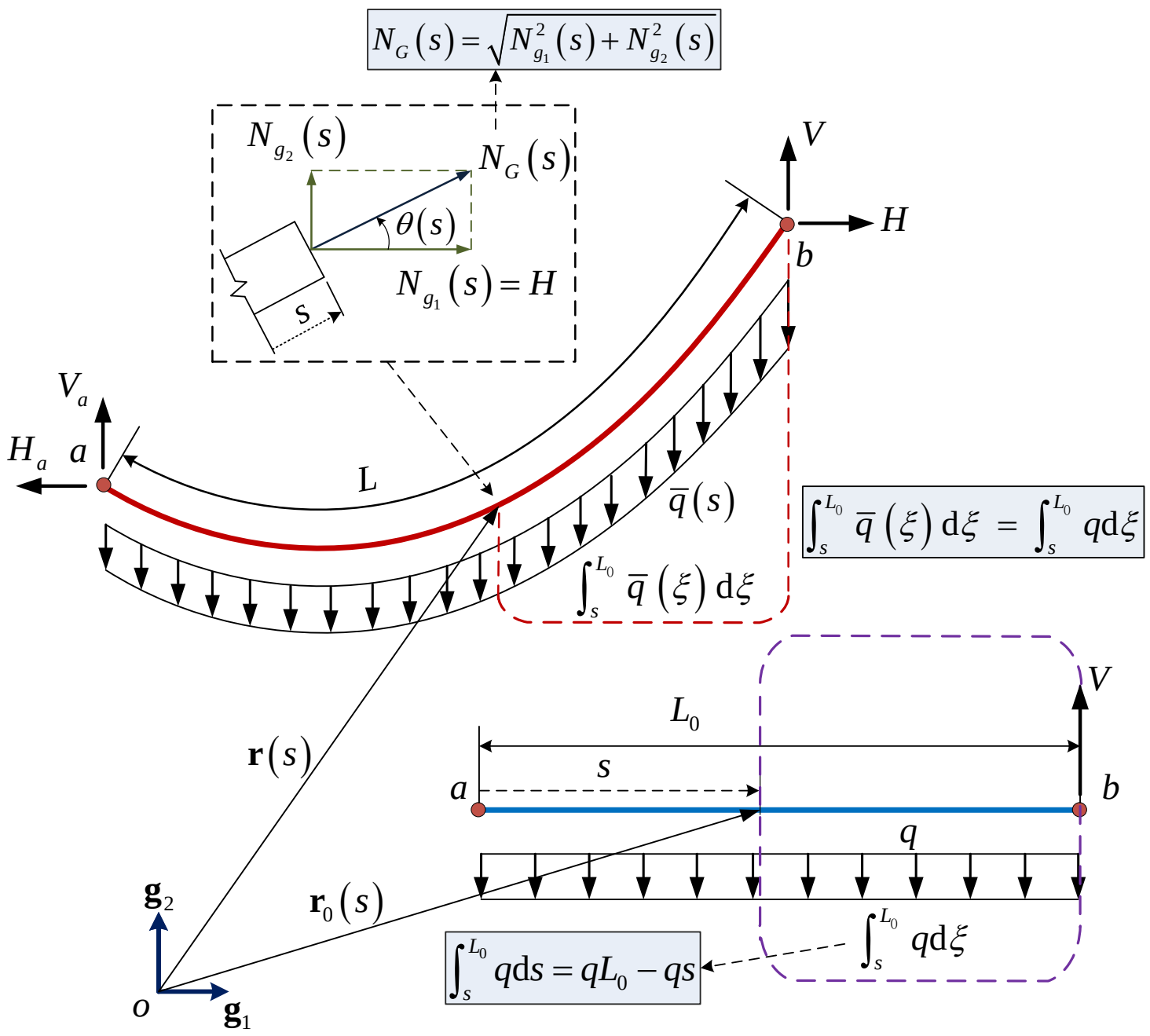

**Fig. 2**. Definition of internal force fields.

Considering that the direction of tension is always perpendicular to the cross-section due to the zero shear deformation as introduced in assumption (A2), the tension $N_G(s)$ $(N_G(s) > 0)$ can be expressed as

$$N_G(s) = \sqrt{N_{g_1}^2(s) + N_{g_2}^2(s)} = \sqrt{H^2 + (V + qL_0 - qs)^2} \tag{17}$$

## 2.4 Description of deformed configuration

Based on assumptions (A1) and (A2), the constitutive equation of the cable's cross-section can be expressed as

$$N_G(s) = C_G(s)\varepsilon_G(s) > 0 \tag{18}$$

where $C_G(s)$ represents the tension stiffness of the cross-section, which is a function of *s*. By substituting Eq. (18), Eq. (14) can be further expressed as

$$r_{1,s}^2(s) + r_{2,s}^2(s) = \left[C_G^{-1}(s)N_G(s)+1\right]^2 \tag{19}$$

By utilizing the relation of Eq. (11) that $r_{2,s}^2(s)N_{g_1}^2(s) = r_{1,s}^2(s)N_{g_2}^2(s)$ into Eq. (19) and considering the relation in Eq. (17), $r_{1,s}^2(s)$ and $r_{2,s}^2(s)$ can be expressed by the stress resultants as

$$r_{1,s}^2(s) = N_{g_1}^2(s)\left[C_G^{-1}(s)+N_G^{-1}(s)\right]^2 \tag{20}$$

$$r_{2,s}^2(s) = N_{g_2}^2(s)\left[C_G^{-1}(s)+N_G^{-1}(s)\right]^2 \tag{21}$$

Taking into account the force and deformation characteristics of the cable, it is observed that the horizontal internal force component and the first-order derivative of the horizontal component of the position vector are consistently positive in the deformed state, namely $N_{g_1}(s) > 0$ and $r_{1,s}(s) > 0$, and it can be inferred that $r_{2,s}(s)N_{g_2}(s) > 0$ according to Eq. (11). Therefore, the following expressions of $r_{1,s}(s)$ and $r_{2,s}(s)$ can be established

$$r_{1,s}(s) = N_{g_1}(s)\left[C_G^{-1}(s)+N_G^{-1}(s)\right] \tag{22}$$

$$r_{2,s}(s) = N_{g_2}(s)\left[C_G^{-1}(s)+N_G^{-1}(s)\right] \tag{23}$$

By integrating $r_{1,s}(s)$ and $r_{2,s}(s)$, the configuration of the cable under a given position at the starting node can be described as follows

$$r_1(s) = r_1^a + \int_0^s r_{1,s}(\xi)\mathrm{d}\xi = r_1^a + \int_0^s N_{g_1}(\xi)\left[C_G^{-1}(\xi)+N_G^{-1}(\xi)\right]\mathrm{d}\xi \tag{24}$$

$$r_2(s) = r_2^a + \int_0^s r_{2,s}(\xi)\mathrm{d}\xi = r_2^a + \int_0^s N_{g_2}(\xi)\left[C_G^{-1}(\xi)+N_G^{-1}(\xi)\right]\mathrm{d}\xi \tag{25}$$

where $r_1^a$ and $r_2^a$ represent the position components at the starting node *a*. It can be observed that the deformed configuration of the cable depends on the internal force fields $N_{g_1}(s)$ and $N_{g_2}(s)$, the tension stiffness $C_G(s)$ and the position of the starting node ($r_1^a$ and $r_2^a$). For the sake of simplicity, the configuration of the cable can be expressed as

$$\mathbf{r}(s) = \mathbf{r}^a + \int_0^s \mathbf{F}_g(\xi)\left[C_G^{-1}(\xi)+N_G^{-1}(\xi)\right]\mathrm{d}\xi \tag{26}$$

where

$$\mathbf{r}^a = \begin{Bmatrix} r_1^a \\ r_2^a \end{Bmatrix},\ \mathbf{F}_g(s) = \begin{Bmatrix} N_{g_1}(s) \\ N_{g_2}(s) \end{Bmatrix} = \begin{Bmatrix} H \\ V + qL_0 - qs \end{Bmatrix} \tag{27}$$

## 3 Implementation of finite element

This section presents the implementation of the solution for cable structures. The cable element is developed based on the exact definition of the tension field, as depicted in Eqs. (15) and (16). The equations of a cable element are established by considering two types of conditions: the equilibrium conditions at element boundaries and the compatibility condition within the element. Subsequently, the linearization of the element equations and the expression of element tangent matrix are provided. Finally, the solution methods for scenarios involving given unstrained length and unstrained length to be solved are presented, respectively.

### 3.1 Equations of a cable element

As indicated by the expressions of the configuration states (Eqs. (24) and (25)) and the internal forces (Eqs. (15) and (16)), it is evident that the element state is wholly defined by the positional quantities at the starting and ending nodes (four quantities), the internal forces at the ending node (the two internal force parameters), and the unstrained length of the cable element. Without additional specified conditions, six equations can be established to construct the equation system of the cable element. **Fig. 3** illustrates a schematic diagram of the established element equation, and the composition of the equation system is detailed below.

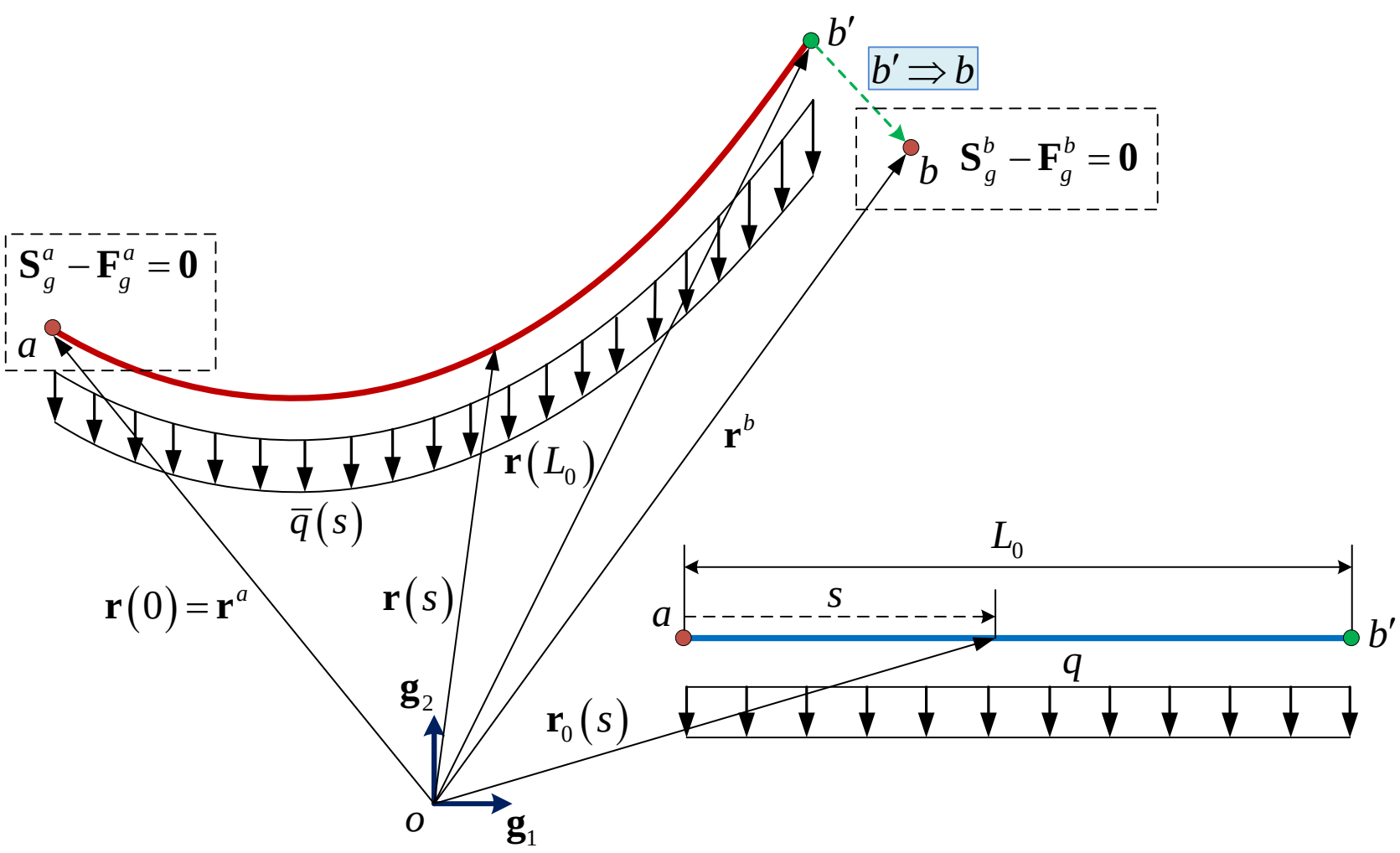

**Fig. 3**. Deformation compatibility of the cable element.

Four equations are obtained according to the boundary conditions of internal forces at the two boundaries, $s = 0$ and $s = L_0$, and they are

$$\mathbf{F}_g^a = -\mathbf{F}_g(0) = -\{H \quad V + qL_0\}^{\mathrm{T}} = \mathbf{S}_g^a \tag{28}$$

$$\mathbf{F}_g^b = \mathbf{F}_g(L_0) = \{H \quad V\}^{\mathrm{T}} = \mathbf{S}_g^b \tag{29}$$

where $\mathbf{S}_g^a$ and $\mathbf{S}_g^b$ can be considered as the external boundary force vectors at the starting and ending nodes, respectively.

Furthermore, it is essential to ensure the compatibility condition within the element. For the proposed cable elements, the kinematical boundary conditions at $s = L_0$ should be satisfied. This implies that the position of the ending node obtained by Eq (26) (point $b'$ in **Fig. 3**) should align with the output position of the ending node (point $b$ in **Fig. 3**). Consequently, the following equation can be formulated.

$$\mathbf{h}_0 = \mathbf{r}^b - \mathbf{r}^a - \int_0^{L_0} \mathbf{r}_{,s}(s)\mathrm{d}s = \mathbf{0} \tag{30}$$

### 3.2 Linearization of the equations

In the context of the finite element method, the implementation of an incremental/iterative solution necessitates the derivation of linearized equations. This section provides a detailed presentation of the linearization of the element equations as outlined in Eqs. (28), (29) and (30) for the unknowns $r_1^a$, $r_2^a$, $r_1^b$, $r_2^b$, $H$, $V$ and $L_0$.

(1) The variation of $\mathbf{F}_g^a$

The variation of $\mathbf{F}_g^a$ in Eq. (28) can be expressed as

$$\delta\mathbf{F}_g^a = -\mathbf{g}_1\delta H - \mathbf{g}_2\delta V - q\mathbf{g}_2\delta L_0 \tag{31}$$

(2) The variation of $\mathbf{F}_g^b$

The variation of $\mathbf{F}_g^b$ in Eq. (29) can be expressed as

$$\delta\mathbf{F}_g^b = \mathbf{g}_1\delta H + \mathbf{g}_2\delta V \tag{32}$$

(3) The variation of $\mathbf{h}_0$

The variation of $\mathbf{h}_0$ in Eq. (30) can be expressed as

$$\delta\mathbf{h}_0 = \mathbf{g}_1\delta r_1^b + \mathbf{g}_2\delta r_2^b - \mathbf{g}_1\delta r_1^a - \mathbf{g}_2\delta r_2^a - \mathbf{B}_H\delta H - \mathbf{B}_V\delta V - \mathbf{B}_L\delta L_0 \tag{33}$$

In Eq. (33), $\mathbf{B}_H$ and $\mathbf{B}_V$ can be obtained by

$$\mathbf{B}_H = \frac{\partial\int_0^{L_0}\mathbf{r}_{,s}(s)\mathrm{d}s}{\partial H} = \int_0^{L_0}\begin{bmatrix}\partial r_{1,s}(s)/\partial H \\ \partial r_{2,s}(s)/\partial H\end{bmatrix}\mathrm{d}s \tag{34}$$

$$\mathbf{B}_V = \frac{\partial\int_0^{L_0}\mathbf{r}_{,s}(s)\mathrm{d}s}{\partial V} = \int_0^{L_0}\begin{bmatrix}\partial r_{1,s}(s)/\partial V \\ \partial r_{2,s}(s)/\partial V\end{bmatrix}\mathrm{d}s \tag{35}$$

where $\partial r_{1,s}(s)/\partial H$, $\partial r_{2,s}(s)/\partial H$, $\partial r_{1,s}(s)/\partial V$ and $\partial r_{2,s}(s)/\partial V$ can be expressed by introducing Eqs. (22), (23), (15) and (16) as

$$\partial r_{1,s}(s)/\partial H = \frac{\partial}{\partial H}\left\{H\left[C_G^{-1}(s) + N_G^{-1}(s)\right]\right\} = C_G^{-1}(s) + N_G^{-1}(s) - H^2 N_G^{-3}(s) \tag{36}$$

$$\partial r_{2,s}(s)/\partial H = \frac{\partial}{\partial H}\left\{(V + qL_0 - qs)\left[C_G^{-1}(s) + N_G^{-1}(s)\right]\right\} = -H(V + qL_0 - qs)N_G^{-3}(s) \tag{37}$$

$$\partial r_{1,s}(s)/\partial V = \frac{\partial}{\partial V}\left\{H\left[C_G^{-1}(s) + N_G^{-1}(s)\right]\right\} = -H(V + qL_0 - qs)N_G^{-3}(s) \tag{38}$$

$$\partial r_{2,s}(s)/\partial V = \frac{\partial}{\partial V}\left\{(V + qL_0 - qs)\left[C_G^{-1}(s) + N_G^{-1}(s)\right]\right\} = C_G^{-1}(s) + N_G^{-1}(s) - (V + qL_0 - qs)^2 N_G^{-3}(s) \tag{39}$$

Furthermore, $\mathbf{B}_L$ in Eq. (33) can be obtained as

$$\mathbf{B}_L = \frac{\partial \int_0^{L_0} \mathbf{r}_{,s}(s)\mathrm{d}s}{\partial L_0} \tag{40}$$

In contrast to the derivation of $\mathbf{B}_H$ and $\mathbf{B}_V$, the derivation of $\mathbf{B}_L$ requires differentiation of the integral limit. Due to the fact that the cross-section stiffness of the cable is a function of *s*, it is difficult to obtain an explicit expression for $\int_0^{L_0} \mathbf{r}_{,s}(s)\mathrm{d}s$, which will make it difficult to analytically derive the expression for $\mathbf{B}_L$. In this study, $\mathbf{B}_L$ is firstly represented as the weighted sum of the function $\mathbf{r}_{,s}(s)$ under a series of *s* values, where the specific values of *s* and the corresponding weights can be determined according to the value of $L_0$ and numerical integration rules. Then, the expression of $\mathbf{B}_L$ can be successfully derived based on the weighted sum expression.

By introducing the $N_{GP}$ integration point $\eta_i \in [0,1] (i = 1,2,\ldots,N_{GP})$ and their weight coefficients $w_i (i = 1,2,\ldots,N_{GP})$ with $\sum_{i=1}^{N_{GP}} w_i = 1$ to implement the numerical integration, $\mathbf{B}_L$ can be expressed as

$$\mathbf{B}_L = \frac{\partial \int_0^{L_0} \mathbf{r}_{,s}(s)\mathrm{d}s}{\partial L_0} = \sum_{i=1}^{N_{GP}} \frac{\partial \left[ (w_i L_0) \mathbf{r}_{,s}(\eta_i L_0) \right]}{\partial L_0} = \sum_{i=1}^{N_{GP}} \begin{bmatrix} w_i r_{1,s}(\eta_i L_0) + (w_i L_0) \partial r_{1,s}(\eta_i L_0) / \partial L_0 \\ w_i r_{2,s}(\eta_i L_0) + (w_i L_0) \partial r_{2,s}(\eta_i L_0) / \partial L_0 \end{bmatrix} \tag{41}$$

By introducing Eqs. (22), (23), (15) and (16), $\partial r_{1,s}(s)/\partial L_0$ and $\partial r_{2,s}(s)/\partial L_0$ can be expressed as

$$\partial r_{1,s}(\eta_i L_0)/\partial L_0 = -q\alpha_i H (V + qL_0\alpha_i) N_G^{-3}(\eta_i L_0) \tag{42}$$

$$\partial r_{2,s}(\eta_i L_0)/\partial L_0 = q\alpha_i C_G^{-1}(\eta_i L_0) + q\alpha_i N_G^{-1}(\eta_i L_0) + q\alpha_i (V + qL_0\alpha_i)^2 N_G^{-3}(\eta_i L_0) \tag{43}$$

where $\alpha_i = 1 - \eta_i$.

To maintain consistency in expression, $\mathbf{B}_H$ and $\mathbf{B}_V$ are also expressed in numerical integration form as

$$\mathbf{B}_H = \int_0^{L_0} \begin{bmatrix} \partial r_{1,s}(s)/\partial H \\ \partial r_{2,s}(s)/\partial H \end{bmatrix} \mathrm{d}s = \sum_{i=1}^{N_{GP}} \begin{bmatrix} (w_i L_0) \partial r_{1,s}(\eta_i L_0)/\partial H \\ (w_i L_0) \partial r_{2,s}(\eta_i L_0)/\partial H \end{bmatrix} \tag{44}$$

$$\mathbf{B}_V = \int_0^{L_0} \begin{bmatrix} \partial r_{1,s}(s)/\partial V \\ \partial r_{2,s}(s)/\partial V \end{bmatrix} \mathrm{d}s = \sum_{i=1}^{N_{GP}} \begin{bmatrix} (w_i L_0) \partial r_{1,s}(\eta_i L_0)/\partial V \\ (w_i L_0) \partial r_{2,s}(\eta_i L_0)/\partial V \end{bmatrix} \tag{45}$$

(4) Incremental equations

Based on the expressions of $\delta \mathbf{F}_g^a$, $\delta \mathbf{F}_g^b$ and $\delta \mathbf{h}_0$, the Taylor series expansion of the element equations for the $(i+1)^{\text{th}}$ step in incremental/iterative solution can be expressed as follows

$$\mathbf{F}_g^{a,i+1} \approx \mathbf{F}_g^{a,i} - \mathbf{g}_1 \Delta H - \mathbf{g}_2 \Delta V - q\mathbf{g}_2 \Delta L_0 = \mathbf{S}_g^{a,i+1} \tag{46}$$

$$\mathbf{F}_g^{b,i+1} \approx \mathbf{F}_g^{b,i} + \mathbf{g}_1 \Delta H + \mathbf{g}_2 \Delta V = \mathbf{S}_g^{b,i+1} \tag{47}$$

$$\mathbf{h}_0^{i+1} \approx \mathbf{h}_0^i - \mathbf{g}_1 \Delta r_1^a - \mathbf{g}_2 \Delta r_2^a + \mathbf{g}_1 \Delta r_1^b + \mathbf{g}_2 \Delta r_2^b - \mathbf{B}_H^i \Delta H - \mathbf{B}_V^i \Delta V - \mathbf{B}_L^i \Delta L_0 = \mathbf{0} \tag{48}$$

where the quantities in current state are denoted as superscript *i*, and $\mathbf{B}_H^i$, $\mathbf{B}_V^i$ and $\mathbf{B}_L^i$ are used to denote $\mathbf{B}_H$, $\mathbf{B}_V$ and $\mathbf{B}_L$ in current state that

$$\mathbf{B}_H^i = \mathbf{B}_H\left(H^i, V^i, L_0^i\right), \mathbf{B}_V^i = \mathbf{B}_V\left(H^i, V^i, L_0^i\right), \mathbf{B}_{L_0}^i = \mathbf{B}_{L_0}\left(H^i, V^i, L_0^i\right) \tag{49}$$

Based on Eqs. (46)-(48), six incremental equations can be established as follows.

$$\begin{bmatrix} \mathbf{0}_{4\times4} & \mathbf{K}_{F_4-\beta_3}^e \\ \mathbf{K}_{h_2-r}^e & \mathbf{K}_{h_2-\beta_3}^{e,i} \end{bmatrix} \begin{Bmatrix} \Delta\mathbf{d}_r^e \\ \Delta\mathbf{d}_{\beta_3}^e \end{Bmatrix} = \begin{Bmatrix} \mathbf{E}_{F_4}^{e,i} \\ \mathbf{E}_{h_2}^{e,i} \end{Bmatrix} \tag{50}$$

where $\mathbf{K}_{F_4-\beta_3}^{e,i}$, $\mathbf{K}_{h_2-r}^{e,i}$ and $\mathbf{K}_{h_2-\beta_3}^{e,i}$ represent the element tangent matrices, $\Delta\mathbf{d}_r^e$ and $\Delta\mathbf{d}_{\beta_3}^e$ represent vectors of incremental element state, $\mathbf{E}_{F_4}^{e,i}$ and $\mathbf{E}_{h_2}^{e,i}$ refer to the element residual force vector and element residual vector for deformation compatibility, respectively, they are expressed as

$$\Delta\mathbf{d}_r^e = \left\{\Delta r_1^a \quad \Delta r_2^a \quad \Delta r_1^b \quad \Delta r_2^b\right\}^{\mathrm{T}} \tag{51}$$

$$\Delta\mathbf{d}_{\beta_3}^e = \left\{\Delta H \quad \Delta V \quad \Delta L_0\right\}^{\mathrm{T}} \tag{52}$$

$$\mathbf{E}_{F_4}^{e,i} = \begin{Bmatrix} \mathbf{S}_g^{a,i+1} - \mathbf{F}_g^{a,i} \\ \mathbf{S}_g^{b,i+1} - \mathbf{F}_g^{b,i} \end{Bmatrix} \tag{53}$$

$$\mathbf{E}_{h_2}^{e,i} = -\mathbf{h}_0^i \tag{54}$$

$$\mathbf{K}_{F_4-\beta_3}^e = \begin{bmatrix} -\mathbf{g}_1 & -\mathbf{g}_2 & -q\mathbf{g}_2 \\ \mathbf{g}_1 & \mathbf{g}_2 & \mathbf{0}_{2\times1} \end{bmatrix} \tag{55}$$

$$\mathbf{K}_{h_2-r}^e = \begin{bmatrix} -\mathbf{g}_1 & -\mathbf{g}_2 & \mathbf{g}_1 & \mathbf{g}_2 \end{bmatrix} \tag{56}$$

$$\mathbf{K}_{h_2-\beta_3}^{e,i} = \begin{bmatrix} -\mathbf{B}_H^i & -\mathbf{B}_V^i & -\mathbf{B}_L^i \end{bmatrix} \tag{57}$$

Within a cable element, there are seven unknowns to be resolved, while only six equations are established in Eq. (50). Hence, an additional condition must be provided for the implementation of the solution. Subsequently, the following section introduces the incremental equation systems of the element for two scenarios: solution with given unstrained length and solution with unstrained length to be solved. **Fig. 4** illustrates an example containing a cable element with given $L_0$ (cable element (1)) and a cable element with unstrained length to be solved under given $N_G\left(L_0\right)$ (cable element (2)).

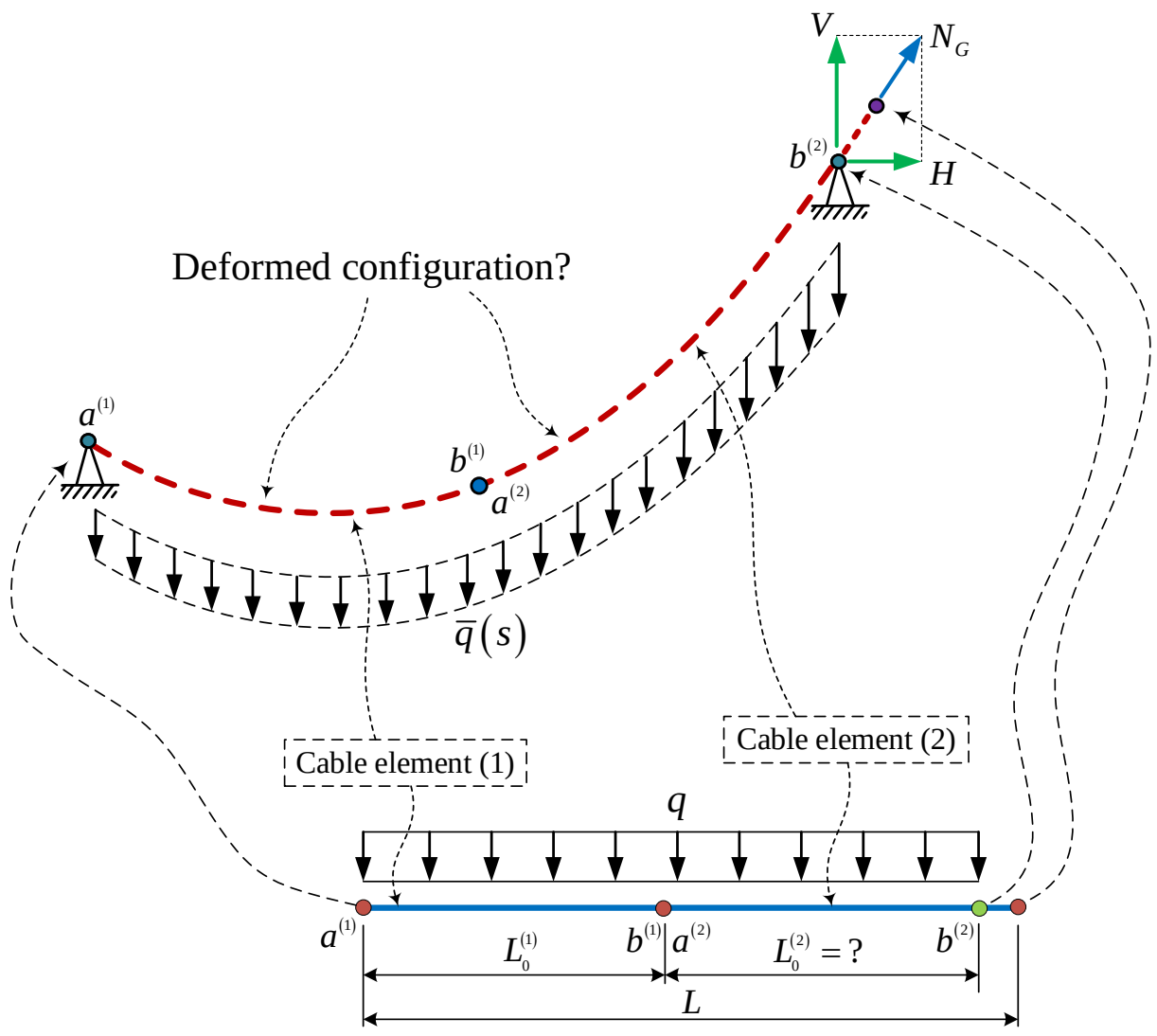

**Fig. 4**. An example with two forms of cable element.

a) Cable element with given unstrained length

For the case where the unstrained length is given, the element incremental equation system can be simplified as follows considering that $L_0$ is a known quantity

$$\begin{bmatrix} \mathbf{0}_{4\times4} & \mathbf{K}^{e}_{F_4-\beta_2} \\ \mathbf{K}^{e}_{h_2-r} & \mathbf{K}^{e,i}_{h_2-\beta_2} \end{bmatrix} \begin{Bmatrix} \Delta\mathbf{d}^{e}_{r} \\ \Delta\mathbf{d}^{e}_{\beta_2} \end{Bmatrix} = \begin{Bmatrix} \mathbf{E}^{e,i}_{F_4} \\ \mathbf{E}^{e,i}_{h_2} \end{Bmatrix} \tag{58}$$

where

$$\Delta\mathbf{d}^{e}_{\beta_2} = \{\Delta H \quad \Delta V\}^{\mathrm{T}} \tag{59}$$

$$\mathbf{K}^{e}_{F_4-\beta_2} = \begin{bmatrix} -\mathbf{g}_1 & -\mathbf{g}_2 \\ \mathbf{g}_1 & \mathbf{g}_2 \end{bmatrix} \tag{60}$$

$$\mathbf{K}^{e,i}_{h_2-\beta_2} = \begin{bmatrix} -\mathbf{B}^{i}_{H} & -\mathbf{B}^{i}_{V} \end{bmatrix} \tag{61}$$

b) Cable element with unstrained length to be solved

For the scenario where the unstrained length is considered as an unknown, an additional equation reflecting an extra relationship or constraint condition must be introduced for the implementation of the solution. The general expression for the additional equation can be written as

$$h_G = G\left(H, V, L_0, r_1^a, r_2^a, r_1^b, r_2^b\right) = 0 \tag{62}$$

Note that Eq. (62) represents the general expression for the constraint equation. The specific constraint equation should be determined based on the specific constraint condition. In the solution, the corresponding constraint is implemented by substituting the specific constraint equation, some examples of which will be provided in the part of numerical examples.

The variation of $h_R$ can be expressed as

$$\delta h_G = \frac{\partial G}{\partial r_1^a}\delta r_1^a + \frac{\partial G}{\partial r_2^a}\delta r_2^a + \frac{\partial G}{\partial r_1^b}\delta r_1^b + \frac{\partial G}{\partial r_2^b}\delta r_2^b + \frac{\partial G}{\partial H}\delta H + \frac{\partial G}{\partial V}\delta V + \frac{\partial G}{\partial L_0}\delta L_0 \tag{63}$$

Then, the Taylor series expansion of Eq. (62) for the ($i$+1)th step in incremental/iterative solution can be expressed as follows

$$h_G^{i+1} \approx h_G^{i} + \frac{\partial G}{\partial r_1^a}\Delta r_1^a + \frac{\partial G}{\partial r_2^a}\Delta r_2^a + \frac{\partial G}{\partial r_1^b}\Delta r_1^b + \frac{\partial G}{\partial r_2^b}\Delta r_2^b + \frac{\partial G}{\partial H}\Delta H + \frac{\partial G}{\partial V}\Delta V + \frac{\partial G}{\partial L_0}\Delta L_0 = 0 \tag{64}$$

By integrating Eq. (64) and Eqs. (46)-(48), the element incremental equation system for incremental/iterative solution with an additional equation can be written by

$$\begin{bmatrix} \mathbf{0}_{4\times4} & \mathbf{K}^{e}_{F_4-\beta_3} \\ \mathbf{K}^{e,i}_{h_3-r} & \mathbf{K}^{e,i}_{h_3-\beta_3} \end{bmatrix} \begin{Bmatrix} \Delta\mathbf{d}^{e}_{r} \\ \Delta\mathbf{d}^{e}_{\beta_3} \end{Bmatrix} = \begin{Bmatrix} \mathbf{E}^{e,i}_{F_4} \\ \mathbf{E}^{e,i}_{h_3} \end{Bmatrix} \tag{65}$$

where

$$\mathbf{E}^{e,i}_{h_3} = \begin{Bmatrix} -\mathbf{h}^{i}_{0} \\ -h^{i}_{G} \end{Bmatrix} \tag{66}$$

$$\mathbf{K}_{h_3-r}^{e,i} = \begin{bmatrix} -\mathbf{g}_1 & -\mathbf{g}_2 & \mathbf{g}_1 & \mathbf{g}_2 \\ \dfrac{\partial G}{\partial r_1^a} & \dfrac{\partial G}{\partial r_2^a} & \dfrac{\partial G}{\partial r_1^b} & \dfrac{\partial G}{\partial r_2^b} \end{bmatrix} \tag{67}$$

$$\mathbf{K}_{h_3-\beta_3}^{e,i} = \begin{bmatrix} -\mathbf{B}_H^i & -\mathbf{B}_V^i & -\mathbf{B}_L^i \\ \dfrac{\partial G}{\partial H} & \dfrac{\partial G}{\partial V} & \dfrac{\partial G}{\partial L_0} \end{bmatrix} \tag{68}$$

(5) Notations of the proposed cable elements

For the sake of simplicity, the proposed cable elements and the numerical models established by the corresponding elements are denoted as follows: a) CET-GUL: Cable element based on Exact Tension field with Given Unstrained Length; b) CET-ULS: Cable element based on Exact Tension field with Unstrained Length to be Solved. Especially, the numerical model including both CET-GUL and CET-ULS is denoted as CET.

### 3.3 Implementation of solution

3.3.1 Solution based on complete tangent matrix

The incremental equation system for the entire structure can be constructed by aggregating the incremental equations of all elements, encompassing Eqs. (58) and (65). Leveraging the element equation system and element tangent matrices obtained in **Sec. 3.1** and **Sec. 3.2**, the global residual vector and tangent matrix can be derived through an assembly process. Notably, each column in the tangent matrix should correspond to the unknown variables in the structural system, while each row in the tangent matrix aligns with the equations constituting the structural system. Regarding the incremental position states $\Delta\mathbf{r}^a$ and $\Delta\mathbf{r}^b$, it is crucial to consider the displacement compatibility between adjacent elements. Generally, for two connected cable elements, the components of the incremental position states at their intersection are consistent, and thus are designated as the same unknowns in the global incremental equation system. In contrast to the incremental position states, the incremental internal force parameters $\Delta H$ and $\Delta V$, as well as the increment unstrained length $\Delta L_0$ of each element, are regarded as independent unknowns in global incremental equation system. If the relationship of internal parameters ($H$, $V$ and $L_0$) between elements is not explicitly specified, the total number of degrees of freedom (DOFs) for a structure containing $n_{e,L_0}$ cable element with given unstrained length and $n_{e,N}$ cable element with unstrained length to be solved is $2n_{node} + 2n_{e,L_0} + 3n_{e,N}$, where $n_{node}$ represents the total number of nodes in the structure. Meanwhile, the total number of incremental equations for the structure is also $2n_{node} + 2n_{e,L_0} + 3n_{e,N}$, comprising $2n_{node}$ equilibrium equations at the nodes and $2n_{e,L_0} + 3n_{e,N}$ equations satisfying the deformation compatibility requirement. For the entire cable structure with $N_{\text{elt}}$ cable elements and $N_{\text{node}}$ nodes, the equation system can be expressed as

$$\begin{cases} \sum_{e\in C_{(j)}} \mathbf{F}_{(j)}^{(e)} - \mathbf{F}_{\text{ext}(j)} = 0 & \left(j = 1,2,\ldots,N_{\text{node}}\right) \\ \mathbf{r}^{b(e)} - \mathbf{r}^{a(e)} - \int_0^{L_0^{(e)}} \mathbf{r}_{,s}^{(e)}\left(s\right)\mathrm{d}s = \mathbf{0} & \left(e = 1,2,\ldots,N_{\text{elt}}\right) \end{cases} \tag{69}$$

where the superscript '(*e*)' represents the element index, the subscript '(*j*)' refers to the index of node, $C_{(j)}$ represents the set of elements related to *j*-th node, $\mathbf{F}_{(j)}^{(e)}$ refers to the nodal force of *e*-th element at *j*-th node, and $\mathbf{F}_{\text{ext}(j)}$ is the external force at *j*-th node.

Following Newton's iteration scheme, the incremental equation system of the whole structure expressed as follows is solved at each iteration step $i = 0,1,2,\ldots$

$$\mathbf{K}_T^{g,i}\Delta\mathbf{d}_g = \mathbf{E}_f^{g,i} \tag{70}$$

where $\mathbf{K}_T^{g,i}$ represents the tangent matrix of the entire structure, $\mathbf{E}_f^{g,i}$ is the residual vector of the entire structure and $\Delta\mathbf{d}_g$ refers to the incremental state vector of the entire structure. By solving the above equations, $\Delta\mathbf{d}_g$ can be obtained and the state vector of the entire structure can be updated by adding $\Delta\mathbf{d}_g$ to the previous state vector $\mathbf{d}_g^i$ as $\mathbf{d}_g^{i+1} = \mathbf{d}_g^i + \Delta\mathbf{d}_g$. It should be noted that $\mathbf{E}_f^{g,i}$ can be expressed as

$$\mathbf{E}_f^{g,i} = \begin{Bmatrix} \hat{\mathbf{E}}_f^{g,i} \\ \mathbf{E}_h^{g,i} \end{Bmatrix} \tag{71}$$

where $\hat{\mathbf{E}}_f^{g,i}$ and $\mathbf{E}_h^{g,i}$ represent the residual force vector of the structure and the residual vector of deformation compatibility, respectively, they can be expressed as

$$\mathbf{E}_h^{g,i} = \begin{Bmatrix} \mathbf{r}^{a(1)} + \int_0^{L_0^{(1)}} \mathbf{r}_{,s}^{(1)}\left(s\right)\mathrm{d}s - \mathbf{r}^{b(1)} \\ \mathbf{r}^{a(2)} + \int_0^{L_0^{(2)}} \mathbf{r}_{,s}^{(2)}\left(s\right)\mathrm{d}s - \mathbf{r}^{b(2)} \\ \cdots \\ \mathbf{r}^{a(N_{\text{elt}})} + \int_0^{L_0^{(N_{\text{elt}})}} \mathbf{r}_{,s}^{(N_{\text{elt}})}\left(s\right)\mathrm{d}s - \mathbf{r}^{b(N_{\text{elt}})} \end{Bmatrix} \tag{72}$$

$$\hat{\mathbf{E}}_f^{g,i} = \hat{\mathbf{S}}_f^{g,i} - \hat{\mathbf{F}}_f^{g,i} \tag{73}$$

with

$$\hat{\mathbf{S}}_f^{g,i} = \begin{Bmatrix} \mathbf{F}_{\text{ext}(1)} \\ \mathbf{F}_{\text{ext}(2)} \\ \cdots \\ \mathbf{F}_{\text{ext}(N_{\text{node}})} \end{Bmatrix}, \quad \hat{\mathbf{F}}_f^{g,i} = \begin{Bmatrix} \sum_{e\in C_{(1)}} \mathbf{F}_{(1)}^{(e)} \\ \sum_{e\in C_{(2)}} \mathbf{F}_{(2)}^{(e)} \\ \cdots \\ \sum_{e\in C_{(N_{\text{node}})}} \mathbf{F}_{(N_{\text{node}})}^{(e)} \end{Bmatrix} \tag{74}$$

The equilibrium configuration for a specified geometrically nonlinear analysis problem can be attained using a standard incremental/iterative approach employing the Newton-Raphson method with load control [41]. In respect of implementation, the convergence condition for equilibrium iteration is expressed as

$$\left\|\mathbf{E}_f^{g,i}\right\| < tol^g \tag{75}$$

where $tol^g$ represents a convergence tolerance in structural level iteration.

3.3.2 Solution based on element internal iteration

The previously described solution method, which relies on the complete tangent matrix, results in a larger scale of the tangent matrix and is contingent upon whether the unstrained length of the element is known. Therefore, it is expected to differentiate between internal and external degrees of freedom based on the attributes of unknowns, with the aim of minimizing the size of the overall tangent matrix and regularizing its number of DOFs. The following solution approach involving two-level iterations is developed in this paper.

Given that the incremental internal force parameters ( $\Delta H$ and $\Delta V$ ) and incremental unstrained length ( $\Delta L_0$ ) are independent unknowns for each cable element, and do not have additional constrain conditions between elements, they can be determined through internal iteration according to the element equation system. Consequently, the equilibrium equations at the structural level can be formulated as

$$\hat{\mathbf{F}}_f^{g,i} = \hat{\mathbf{S}}_f^{g,i} \tag{76}$$

where $\hat{\mathbf{F}}_f^{g,i}$ is the nodal force vector of the structure, and $\hat{\mathbf{S}}_f^{g,i}$ is the external nodal load vector of the structure. Then, the incremental equation system for the structure can be expressed as

$$\hat{\mathbf{K}}_T^{g,i}\Delta\hat{\mathbf{d}}_g = \hat{\mathbf{S}}_f^{g,i} - \hat{\mathbf{F}}_f^{g,i} = \hat{\mathbf{E}}_f^{g,i} \tag{77}$$

where $\hat{\mathbf{K}}_T^{g,i}$ represents the tangent stiffness matrix of the structure with the size of $2n_{node} \times 2n_{node}$ , and $\Delta\hat{\mathbf{d}}_g$ refers to the incremental position state vector of the structure made up of the position vector $\Delta\mathbf{d}_r^e$ of all nodes. By solving the above equation, $\Delta\hat{\mathbf{d}}_g$ can be obtained and the position vector of all nodes can be updated by adding $\Delta\hat{\mathbf{d}}_g$ to the previous position vector of the structure $\hat{\mathbf{d}}_g^i$ as $\hat{\mathbf{d}}_g^{i+1} = \hat{\mathbf{d}}_g^i + \Delta\hat{\mathbf{d}}_g$ . Especially, the tangent stiffness matrix of the structure is obtained by assembling operation of the condensed element tangent matrix $\mathbf{K}_{T,c}^{e,i}$ with the size of $4\times 4$ , which is expressed as follows.

a) For cable element with given unstrained length

$$\mathbf{K}_{T,c}^{e,i} = -\mathbf{K}_{F_4-\beta_2}^{e}\left(\mathbf{K}_{h_2-\beta_2}^{e,i}\right)^{-1}\mathbf{K}_{h_2-r}^{e} \tag{78}$$

b) For cable element with unstrained length to be solved

$$\mathbf{K}_{T,c}^{e,i} = -\mathbf{K}_{F_4-\beta_3}^{e}\left(\mathbf{K}_{h_3-\beta_3}^{e,i}\right)^{-1}\mathbf{K}_{h_3-r}^{e} \tag{79}$$

The convergence condition for equilibrium iteration in structural level is expressed as

$$\left\|\hat{\mathbf{E}}_f^{g,i}\right\| < tol^g \tag{80}$$

In the element-level solution, it is essential to iteratively ascertain the internal force parameters and the unstrained length based on the assumption of fixed nodal positions, to guarantee the deformation compatibility of the cable element. In this iterative process within a cable element, the internal force parameters and the unstrained length can be updated through

$$\mathbf{d}_{\beta_2}^{e,j+1} = \mathbf{d}_{\beta_2}^{e,j} + \left(\mathbf{K}_{h_2-\beta_2}^{e,j}\right)^{-1}\mathbf{E}_{h_2}^{e,j} \tag{81}$$

$$\mathbf{d}_{\beta_3}^{e,j+1} = \mathbf{d}_{\beta_3}^{e,j} + \left(\mathbf{K}_{h_3-\beta_3}^{e,j}\right)^{-1}\mathbf{E}_{h_3}^{e,j} \tag{82}$$

for the cable element with given unstrained length and unstrained length to be solved, respectively, where $j = 1,2,3,\ldots$ is the number of iterations. The convergence condition for element-level iteration is expressed as

$$R < tol^e \tag{83}$$

where $tol^e$ is the convergence tolerance in element-level iteration and $R$ is

$$R = \begin{cases} \|\mathbf{h}_0\| / \|\Delta\mathbf{r}\| & \text{for given } L_0 \\ \|\mathbf{h}_0\| / \|\Delta\mathbf{r}\| + \|h_G\| / \|H\| & \text{for } L_0 \text{ to be solved} \end{cases} \tag{84}$$

For a better understanding, the flowchart of structural state determination with element-level iteration under given load conditions is presented in **Fig. 5**.

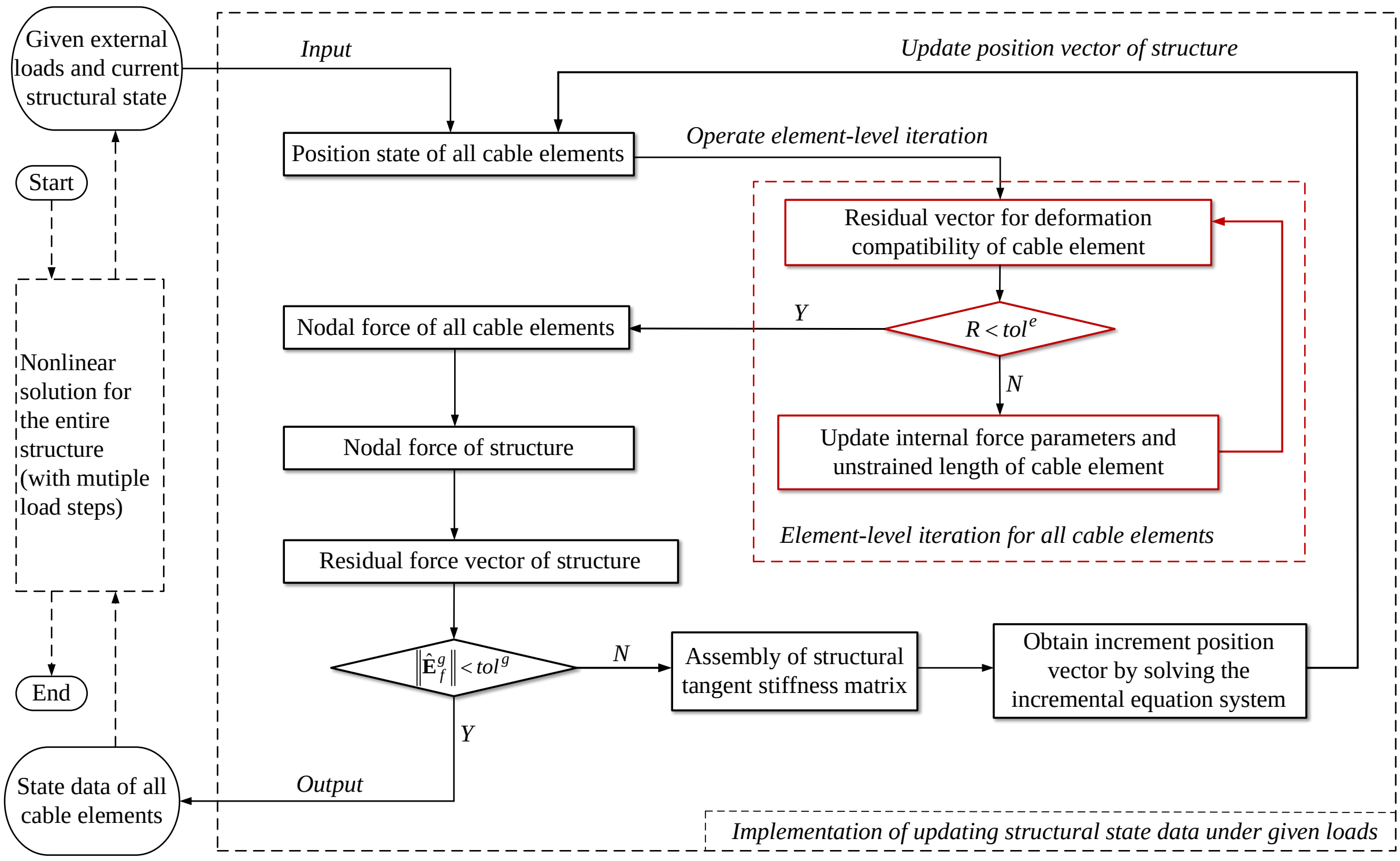

**Fig. 5**. Flowchart of structural state determination under given load.

It is worth emphasizing that while the aforementioned solution method with two-level iterations effectively conserves computational resources and ensures the regularity of the numerical model, its applicability is more restricted compared to the solution method with complete tangent matrix. The limitation of the two-level iterative method primarily lies in its inability to accommodate scenarios where the force parameters and unstressed length have additional constraint relationships between elements, as this method treats force parameters and unstressed length as internal variables of the element and achieves condensation based on the internal conditions of the element. Therefore, for cable structures without additional constraint relationships between elements on force parameters and unstressed length, such as Example 1 and Examples 3-5 in **Sec. 4**, the two-level iterative approach is adopted as the main solution method. Nevertheless, for the cases with the additional constraints, such as Example 2 in **Sec. 4**, the method with complete tangent matrix must be utilized for resolution.

### 3.4 Three-dimensional cable element

Although only the formulation of planar cable elements is presented in **Sec. 3.1** and **Sec. 3.3**, they can be easily extended to three-dimensional cable elements by introducing appropriate transformations. **Fig. 6** demonstrates the relationship between the global coordinate system (*xyz*) and the local coordinate system of a cable element ($\mathbf{g}_1 - \mathbf{g}_2$). As displayed in **Fig. 6**, the base vector $\mathbf{g}_2$ for the cable element is consistent with *z*-axis, and the direction of $\mathbf{g}_1$ can be described by the angle $\varphi$. Then, the nodal forces in global coordinate system can be expressed as

$$\begin{aligned} H_{ax} &= H_a \cos\varphi, \quad H_{bx} = H_b \cos\varphi \\ H_{ay} &= H_a \sin\varphi, \quad H_{by} = H_b \sin\varphi \end{aligned} \tag{85}$$

where $H_{ax}, H_{bx}, H_{ay}, H_{by}$ are the nodal forces corresponding to the global coordinate system (*xyz*), and $H_a, H_b$ refer to the nodal forces corresponding to local coordinate system of the cable element ($\mathbf{g}_1 - \mathbf{g}_2$).

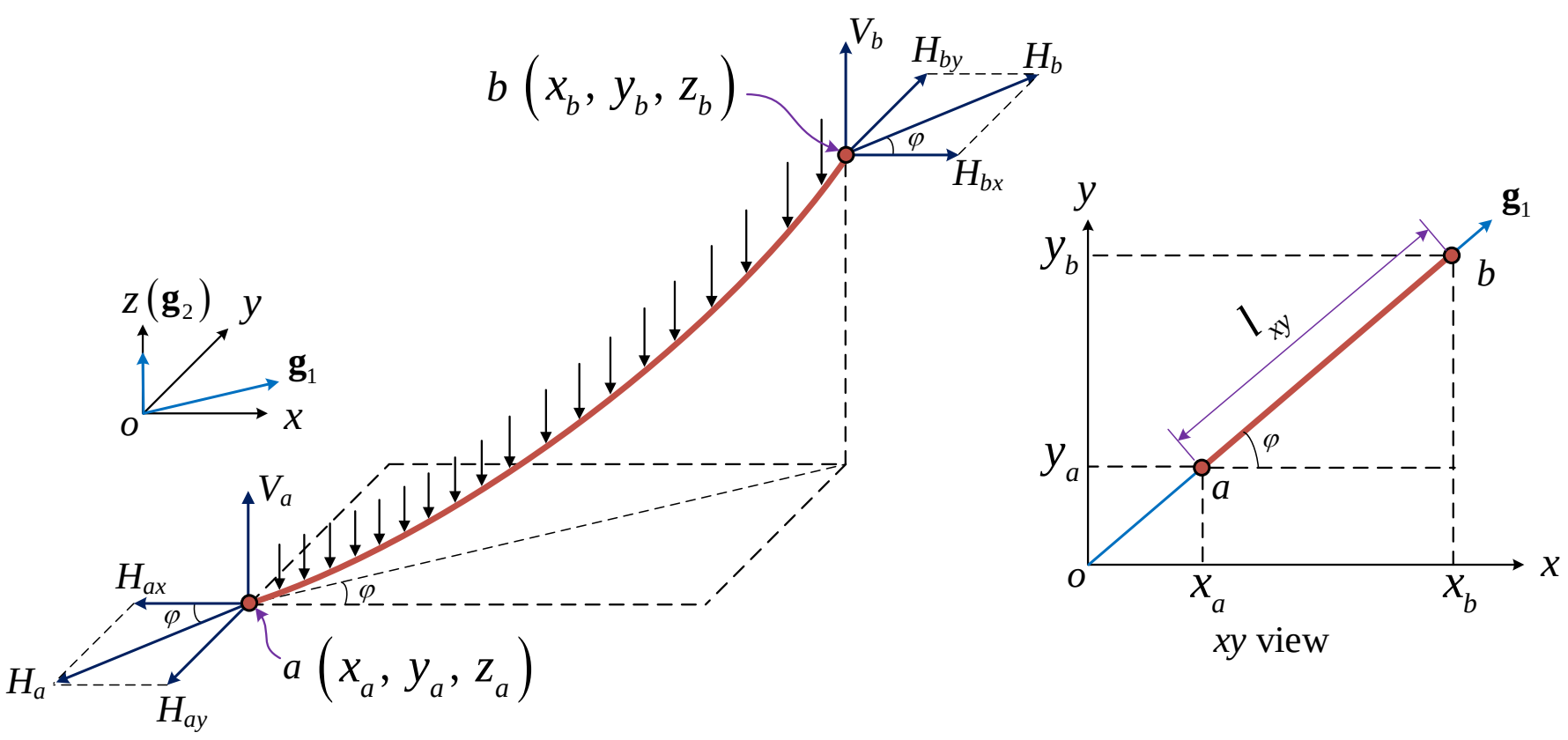

**Fig. 6**. The cable element in global coordinate system.

For the sake of simplicity, the CET-GUL with element internal iteration is taken as an example to explain the extended formulation of the three-dimensional cable elements. The following formulation is used for Examples 3 and 4 in **Sec. 4**.

Based on the above relationship (Eq. (85)), the transformation between global coordinate system and local coordinate system for $e^{\text{th}}$ cable element can be established as

$$\mathbf{F}_{xyz}^{e} = \mathbf{T}^{e} \mathbf{F}_{g}^{e} \tag{86}$$

where the superscript '*e*' represents the index of the element, and $\mathbf{F}_{xyz}^{e}$ and $\mathbf{F}_{g}^{e}$ are

$$\mathbf{F}_{xyz}^{e} = \left\{ H_{ax}^{e} \quad H_{ay}^{e} \quad V_{a}^{e} \quad H_{bx}^{e} \quad H_{by}^{e} \quad V_{b}^{e} \right\}^{\mathrm{T}} \tag{87}$$

$$\mathbf{F}_{g}^{e} = \left\{ H_{a}^{e} \quad V_{a}^{e} \quad H_{b}^{e} \quad V_{b}^{e} \right\}^{\mathrm{T}} \tag{88}$$

In Eq. (86), $\mathbf{T}^{e}$ is the transformation matrix obtained by

$$\mathbf{T}^e=\begin{bmatrix}\cos\varphi & 0 & 0 & 0\\ \sin\varphi & 0 & 0 & 0\\ 0 & 1 & 0 & 0\\ 0 & 0 & \cos\varphi & 0\\ 0 & 0 & \sin\varphi & 0\\ 0 & 0 & 0 & 1\end{bmatrix}=\frac{1}{l_{xy}}\begin{bmatrix}x_b^e-x_a^e & 0 & 0 & 0\\ y_b^e-y_a^e & 0 & 0 & 0\\ 0 & 1 & 0 & 0\\ 0 & 0 & x_b^e-x_a^e & 0\\ 0 & 0 & y_b^e-y_a^e & 0\\ 0 & 0 & 0 & 1\end{bmatrix} \tag{89}$$

with

$$l_{xy}=\sqrt{\left(x_b^e-x_a^e\right)^2+\left(y_b^e-y_a^e\right)^2} \tag{90}$$

In the global coordinate system, the equilibrium equations of the element can be written as

$$\mathbf{F}_{xyz}^e-\mathbf{S}_{xyz}^e=\mathbf{0} \tag{91}$$

where $\mathbf{S}_{xyz}^e$ represents the vector of external boundary forces at the nodes. In order to achieve iterative solution, it is necessary to derive the corresponding incremental equations and tangent stiffness matrix. Firstly, for the current configuration, the variation of nodal positions between global coordinate system and local coordinate system can be expressed as follows based on the relationship in Eq. (86)

$$\delta\mathbf{r}^e=\left(\mathbf{T}^e\right)^{\mathrm{T}}\delta\mathbf{p}^e \tag{92}$$

where $\mathbf{p}^e=\left\{x_a^e \quad y_a^e \quad z_a^e \quad x_b^e \quad y_b^e \quad z_b^e\right\}^{\mathrm{T}}$. Then, the variation of $\mathbf{F}_{xyz}^e$ can be expressed as

$$\delta\mathbf{F}_{xyz}^e=\mathbf{T}^e\delta\mathbf{F}_g^e+\delta\mathbf{T}^e\mathbf{F}_g^e=\mathbf{T}^e\mathbf{K}_{T,c}^e\left(\mathbf{T}^e\right)^{\mathrm{T}}\delta\mathbf{p}^e+\delta\mathbf{T}^e\mathbf{F}_g^e \tag{93}$$

where $\mathbf{K}_{T,c}^e$ is determined by Eq. (78).

In Eq. (93), $\delta\mathbf{T}^e\mathbf{F}_g^e$ can be further expressed as

$$\delta\mathbf{T}^e\mathbf{F}_g^e=\frac{\partial\mathbf{T}^e\mathbf{F}_g^e}{\partial x_a^e}\delta x_a^e+\frac{\partial\mathbf{T}^e\mathbf{F}_g^e}{\partial y_a^e}\delta y_a^e+\frac{\partial\mathbf{T}^e\mathbf{F}_g^e}{\partial y_a^e}\delta z_a^e+\frac{\partial\mathbf{T}^e\mathbf{F}_g^e}{\partial x_b^e}\delta x_b^e+\frac{\partial\mathbf{T}^e\mathbf{F}_g^e}{\partial y_b^e}\delta y_b^e+\frac{\partial\mathbf{T}^e\mathbf{F}_g^e}{\partial y_b^e}\delta z_b^e=\mathbf{K}_{T,G}^e\delta\mathbf{p}^e \tag{94}$$

where $\mathbf{K}_{T,G}^e$ refers to the geometric stiffness matrix of the element and is obtained by

$$\mathbf{K}_{T,G}^e=\begin{bmatrix}H_aC_{x_a} & H_aC_{y_a} & 0 & H_aC_{x_b} & H_aC_{y_b} & 0\\ H_aS_{x_a} & H_aS_{y_a} & 0 & H_aS_{x_b} & H_aS_{y_b} & 0\\ 0 & 0 & 0 & 0 & 0 & 0\\ H_bC_{x_a} & H_bC_{y_a} & 0 & H_bC_{x_b} & H_bC_{y_b} & 0\\ H_bS_{x_a} & H_bS_{y_a} & 0 & H_bS_{x_b} & H_bS_{y_b} & 0\\ 0 & 0 & 0 & 0 & 0 & 0\end{bmatrix} \tag{95}$$

with

$$\begin{aligned}
C_{x_a}&=-l_{xy}^{-1}+\left(x_b^e-x_a^e\right)^2 l_{xy}^{-3},\ C_{y_a}=\left(x_b^e-x_a^e\right)\left(y_b^e-y_a^e\right)l_{xy}^{-3}\\
C_{x_b}&=l_{xy}^{-1}-\left(x_b^e-x_a^e\right)^2 l_{xy}^{-3},\ C_{y_b}=-\left(x_b^e-x_a^e\right)\left(y_b^e-y_a^e\right)l_{xy}^{-3}\\
S_{x_a}&=\left(x_b^e-x_a^e\right)\left(y_b^e-y_a^e\right)l_{xy}^{-3},\ S_{y_a}=-l_{xy}^{-1}+\left(y_b^e-y_a^e\right)^2 l_{xy}^{-3}\\
S_{x_b}&=-\left(x_b^e-x_a^e\right)\left(y_b^e-y_a^e\right)l_{xy}^{-3},\ S_{y_b}=l_{xy}^{-1}-\left(y_b^e-y_a^e\right)^2 l_{xy}^{-3}
\end{aligned} \tag{96}$$

The Taylor series expansion of Eq. (91) for the $(i+1)^{\mathrm{th}}$ step in incremental/iterative solution can be expressed as

$$\mathbf{F}_{xyz}^{e,i+1} \approx \mathbf{F}_{xyz}^{e,i} + \mathbf{K}_T^e \Delta\mathbf{p}^e = \mathbf{S}_{xyz}^{e,i+1} \quad (97)$$

Then, the incremental equations can be established as

$$\mathbf{K}_T^e \Delta\mathbf{p}^e = \mathbf{S}_{xyz}^{e,i+1} - \mathbf{F}_{xyz}^{e,i} \quad (98)$$

and the tangent stiffness matrix of the cable element can be determined as

$$\mathbf{K}_T^e = \mathbf{T}^e \mathbf{K}_{T,c}^e \left(\mathbf{T}^e\right)^{\mathrm{T}} + \mathbf{K}_{T,G}^e \quad (99)$$

## 4 Numerical examples

The performance of the proposed cable elements is evaluated using multiple numerical examples. Key aspects of the implementation and solution algorithm are highlighted as follows: (1) The efficiency of the solution process is gauged by the number of elements needed to reach convergence. (2) Depending on the problem, the final equilibrium states are achieved using an iterative solution method, which employs either a load control or arc-length control strategy. (3) Unless specified otherwise, the convergence tolerance for the global system and individual elements are set to $tol^g = 1.0\times10^{-8}$ and $tol^e = 1.0\times10^{-8}$, respectively.

The proposed cable elements and corresponding solution algorithms provided in **Sec. 3** have been programed in MATLAB and run on a computer having an Intel® CoreTM i7-8700 processor and a CPU at 3.2GHz with 64GB of RAM.

### 4.1 Example 1: a cable under self-weight

The computational performance of the CET-GUL and CET-ULS is validated using an isolated cable, depicted in **Fig. 7**. The cable span is designated as $l_h = 304.8\text{m}$. Three cases of height differences between the two supports as follows are examined

$$l_v = \begin{cases} 0 & \text{for Case A} \\ 50\text{m} & \text{for Case B} \\ 100\text{m} & \text{for Case C} \end{cases} \quad (100)$$

The self-weight per unit unstrained length of the cable is specified as $q = 5.0\,\text{kN/m}$, while the elastic modulus and cross-sectional area are set to $1.310\times10^8\,\text{kN/m}^2$ and $548.4\times10^{-6}\,\text{m}^2$, respectively. The two ends of the cable are denoted as *a* and *b*, respectively, with the midpoint corresponding to the unstrained configuration represented by *c*. Under the self-weight of the cable, two solution problems are addressed: (1) Determination of equilibrium state under given unstrained length, and (2) Determination of equilibrium state under given horizontal force. In implementation, the solution method involving element internal iteration is employed, addressing both scenarios with given unstrained length and given horizontal force, respectively.

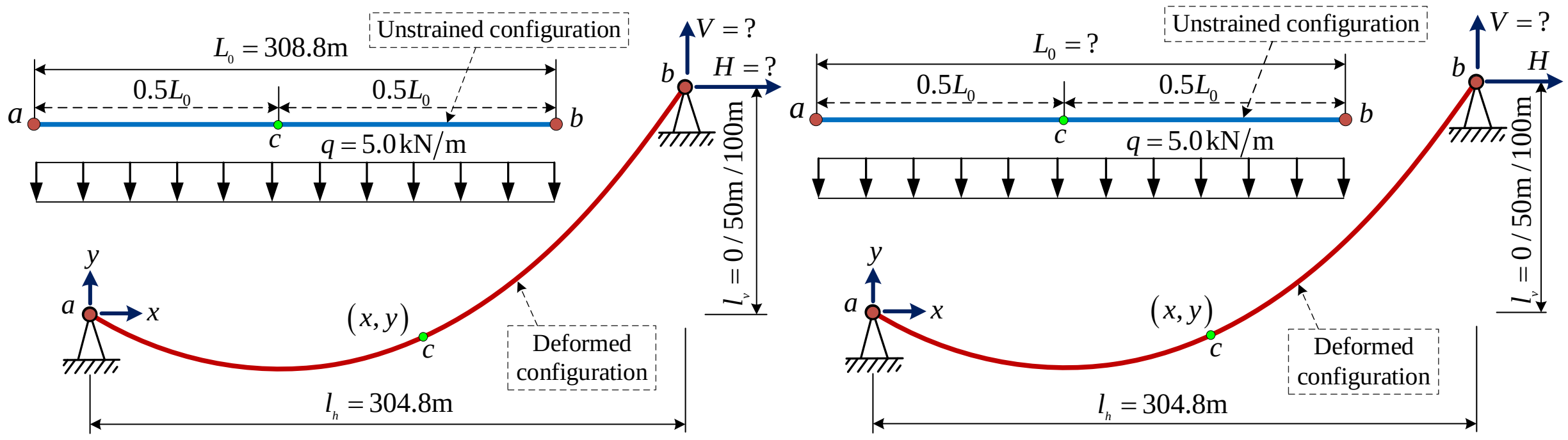

(a) Solution problem under given $L_0$ (b) Solution problem under given $H$

**Fig. 7**. Description of solution problems for the isolated cable.

4.1.1 Solution for the problem with given unstrained length

As depicted in **Fig. 7**a), the unstrained length of the cable is specified as $L_0 = 308.8\text{m}$, and subsequently, the equilibrium state of the cable can be determined using an iterative solution algorithm with a numerical model comprising various elements. For comparative purposes, the other three numerical models are employed to address the problem:

(a) Numerical model utilizing truss element, denoted as TRUSS. In this model, the cable is represented by truss elements, and the self-weight of the cable is modeled as nodal load. To circumvent potential issues related to singular structural stiffness during the initial stages of the solution, the dynamic relaxation method [5] is adopted as the primary solution strategy. Upon reaching a stage where the tension of the truss elements is adequate to avoid structural stiffness singularity, the Newton-Raphson method [41] with static load control is utilized to expedite solution convergence.

(b) Numerical model employing beam element, denoted as BEAM. In this model, the cable is modeled using beam elements constructed based on the rotation-free planar Kirchhoff rod formulation [18], with the self-weight of the cable also represented by nodal load. Notably, instead of NURBS as used in Ref. [18], cubic Hermite interpolation is employed for discretization of the beam element. The Newton-Raphson method [41] with static load control is deployed for solving. Additionally, the flexural stiffness of the cross-section is set to a relatively small value to align the characteristics of the beam element more closely with those of the cable.

(c) Numerical model featuring catenary element based on explicit analytical functions, denoted as CATENARY. In this model, the catenary cable element and solution method are implemented in accordance with the formulation provided by Ref. [26, 27].

The coordinates ($x$, $y$) of point $c$ in the equilibrium states obtained using the four numerical models with varying numbers of equal length elements are presented in **Table 1**, **Table 2** and **Table 3**, respectively, for the three cases involving different height differences between the two supports. Additionally, the horizontal and vertical components of cable force at point $b$ ($H$ and $V$) obtained by TRUSS and CET-GUL are displayed in **Table 4**. In these four tables, $N_e$ represents the number of elements used in the numerical models.

**Table 1**, **Table 2** and **Table 3** demonstrate that CET-GUL and TRUSS yield consistent displacement solutions (with consideration to five significant digits). In scenarios where there is adequate refinement, the numerical model utilizing truss elements can account for the influence of tension and reflect the stiffness characteristics of the cable (without flexural stiffness). Consequently, the convergent results obtained from TRUSS can be deemed accurate and utilized as reference data. For the three considered cases, a single CET-GUL is found to be adequate for achieving convergence and producing precise displacement solutions consistent with the convergent solution provided by TRUSS.

For the BEAM, it is often necessary to assign a relatively small value to the flexural stiffness to closely emulate the mechanical characteristics of the cable. For Case B and Case C, the flexural stiffness of the cross-section is set to $0.01\text{kN/m}^2$, resulting in convergent displacement solutions obtained by BEAM that align with those of TRUSS. However, for Case A, a low value of cross-section flexural stiffness can trigger numerical instability during the solution process. After conducting trial computations, it becomes essential to set the flexural stiffness to at least $1.0\times10^4\ \text{kN/m}^2$ to ensure numerical stability. Owing to the influence of flexural stiffness, a minor disparity exists between the convergent displacement solution of BEAM and that of TRUSS. Additionally, due to the fact that the beam elements utilized in BEAM are based on cubic interpolation functions, which do not precisely replicate the actual cable shape, a substantial number of elements is still necessary for BEAM to achieve convergent displacement solutions.

The computational performance of CATENARY is contingent upon the setting of $l_v$. For Case A, a single catenary element proves adequate for deriving a convergent displacement solution, while for Case B and Case C, multiple catenary elements are requisite for achieving convergent results. From the standpoint of precise convergent displacement solutions, CATENARY cannot generate displacement solutions consistent with those of TRUSS (in terms of five significant digits). This indicates that despite its quicker convergence speed compared to TRUSS and BEAM, errors in the displacement solutions obtained by CATENARY arise from the simplified approximation of the analytical function used to construct the catenary element.

**Table 1** Location (*x*, *y*) of point *c* obtained by the four numerical models (Case A).

| $N_e$ | (*x*, *y*) /m | | | |
|---|---|---|---|---|
| | TRUSS | BEAM | CATENARY | CET-GUL |
| 1 | - | - | <u>152.40, -36.337</u> | <u>152.40, -36.132</u> |
| 2 | 152.40, -40.320 | 152.40, -37.092 | 152.40, -36.337 | 152.40, -36.132 |
| 4 | 152.40, -37.033 | 152.40, -36.120 | | |
| 8 | 152.40, -36.350 | 152.40, -36.141 | | |
| 16 | 152.40, -36.186 | 152.40, -36.137 | | |
| 32 | 152.40, -36.146 | 152.40, -36.134 | | |
| 64 | 152.40, -36.135 | <u>152.40, -36.133</u> | | |
| 128 | 152.40, -36.133 | 152.40, -36.133 | | |
| 256 | <u>152.40, -36.132</u> | | | |
| 512 | 152.40, -36.132 | | | |
| Converged | 152.40, -36.132 | 152.40, -36.133 | 152.40, -36.337 | 152.40, -36.132 |

**Table 2** Location ($x$, $y$) of point $c$ obtained by the four numerical models (Case B).

| $N_e$ | ($x$, $y$) /m | | | |
|---|---|---|---|---|
| | TRUSS | BEAM | CATENARY | CET-GUL |
| 1 | - | - | 157.20, -5.9691 | 157.16, -5.6887 |
| 2 | 157.80, -8.7963 | 157.32, -5.9550 | 157.20, -5.9694 | 157.16, -5.6887 |
| 4 | 157.30, -6.3599 | 157.16, -5.5375 | 157.20, -5.9737 | |
| 8 | 157.19, -5.8511 | 157.16, -5.6580 | 157.20, -5.9745 | |
| 16 | 157.17, -5.7289 | 157.16, -5.6836 | 157.20, -5.9747 | |
| 32 | 157.16, -5.6987 | 157.16, -5.6878 | 157.20, -5.9747 | |
| 64 | 157.16, -5.6911 | 157.16, -5.6885 | | |
| 128 | 157.16, -5.6893 | 157.16, -5.6887 | | |
| 256 | 157.16, -5.6889 | 157.16, -5.6887 | | |
| 512 | 157.16, -5.6888 | | | |
| 1024 | 157.16, -5.6887 | | | |
| 2048 | 157.16, -5.6887 | | | |
| Converged | 157.16, -5.6887 | 157.16, -5.6887 | 157.20, -5.9747 | 157.16, -5.6887 |

**Table 3** Location ($x$, $y$) of point $c$ obtained by the four numerical models (Case C).

| $N_e$ | ($x$, $y$) /m | | | |
|---|---|---|---|---|
| | TRUSS | BEAM | CATENARY | CET-GUL |
| 1 | - | - | 157.73, 32.687 | 157.57, 33.277 |
| 2 | 154.84, 41.724 | 157.24, 34.458 | 157.73, 32.685 | 157.57, 33.277 |
| 4 | 155.79, 38.786 | 157.44, 33.719 | 157.73, 32.683 | |
| 8 | 156.79, 35.671 | 157.53, 33.391 | 157.73, 32.683 | |
| 16 | 157.42, 33.719 | 157.56, 33.305 | | |
| 32 | 157.56, 33.290 | 157.56, 33.284 | | |
| 64 | 157.57, 33.277 | 157.57, 33.279 | | |
| 128 | 157.57, 33.277 | 157.57, 33.278 | | |
| 256 | | 157.57, 33.277 | | |
| 512 | | 157.57, 33.277 | | |
| Converged | 157.57, 33.277 | 157.57, 33.277 | 157.73, 32.683 | 157.57, 33.277 |

As depicted in **Table 4**, CET-GUL achieves convergent force solutions using only one element for the three cases with different $l_v$. In contrast, TRUSS necessitates more elements to attain convergent force solutions compared to achieving convergent displacement solutions. From the outcomes presented in **Table 4**, it is evident that as the number of elements increases, the force solutions obtained by TRUSS progressively approach the convergent solutions obtained by CET-GUL, underscoring the exceptional accuracy of CET-GUL in terms of solution accuracy for internal forces.

**Table 4** The components of cable force at point *b* obtained by TRUSS and CET.

| Case | TRUSS | | | CET-GUL | | |
|---|---|---|---|---|---|---|
| | $N_e$ | $H$/kN | $V$/kN | $N_e$ | $H$/kN | $V$/kN |
| A | 256 | 1599.96 | 768.984 | 1 | 1599.97 | 772.000 |
| | 512 | 1599.96 | 770.489 | 2 | 1599.97 | 772.000 |
| | 1024 | 1599.97 | 771.245 | | | |
| | 2048 | 1599.97 | 771.623 | | | |
| | 4092 | 1599.97 | 771.811 | | | |
| B | 1024 | 1844.57 | 1089.55 | 1 | 1844.57 | 1090.30 |
| | 2048 | 1844.57 | 1089.92 | 2 | 1844.57 | 1090.30 |
| | 4092 | 1844.57 | 1090.11 | | | |
| | 8184 | 1844.57 | 1090.21 | | | |
| C | 64 | 3179.68 | 1820.46 | 1 | 3179.78 | 1832.56 |
| | 128 | 3179.76 | 1826.52 | 2 | 3179.78 | 1832.56 |
| | 256 | 3179.78 | 1829.55 | | | |
| | 512 | 3179.78 | 1831.06 | | | |

Additionally, the solution process of CET-GUL with two elements is examined, and the first two update processes for Case B are illustrated in **Fig. 8**. Typically, each update process comprises two steps. In step 1, the element state determination is carried out through element-level iteration starting from the initial state of the element, where the determined state of each element in the previous update process is used as the initial state. For the first update process, the initial state of each element is set by the user. In step 2, the incremental positions of all nodes are determined via global solution based on the established state of all elements. **Fig. 8**a) illustrates the first update process, wherein the initial position of *c* is set to the midpoint of *a* and *b*, while the state of the element is placed horizontally with an unstrained length. As shown in **Fig. 8**a), the shape of each element undergoes changes during the element state determination process. Typically, significant changes occur in the initial iterations, but as the iterations progress, convergence is gradually achieved. **Fig. 8**b) showcases the second update process, where the elements steadily converge to the state related to the updated nodal positions determined in the previous update process. In practical calculations, the number of iterations in element state determination during the first and second update processes does not exceed 9 and 6, respectively. Subsequently, in the solving process, the number of iterations in element state determination gradually diminishes. At the global level, only six iterations are required to complete the solution, affirming the outstanding convergence of the solution process.

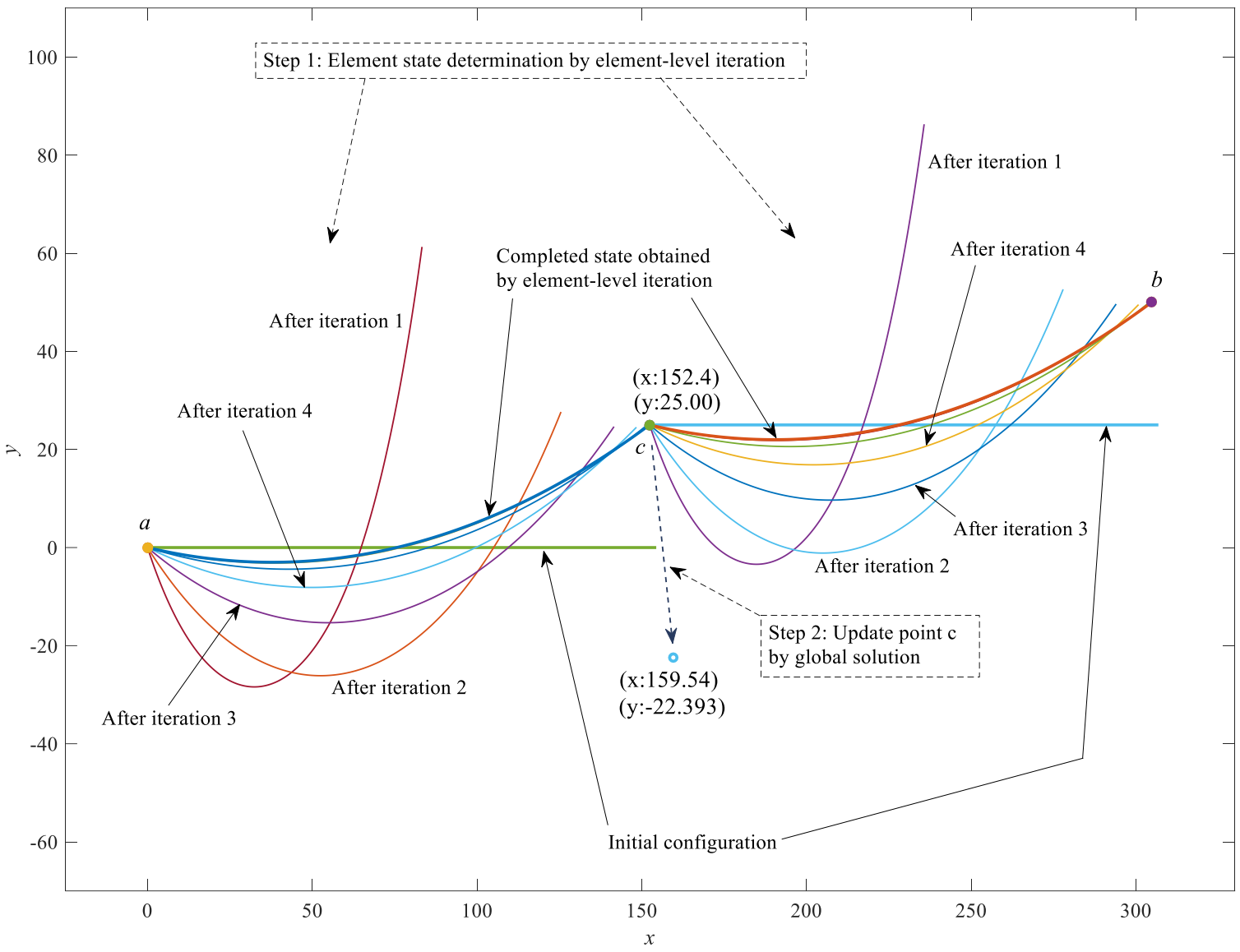

(a) The first update process

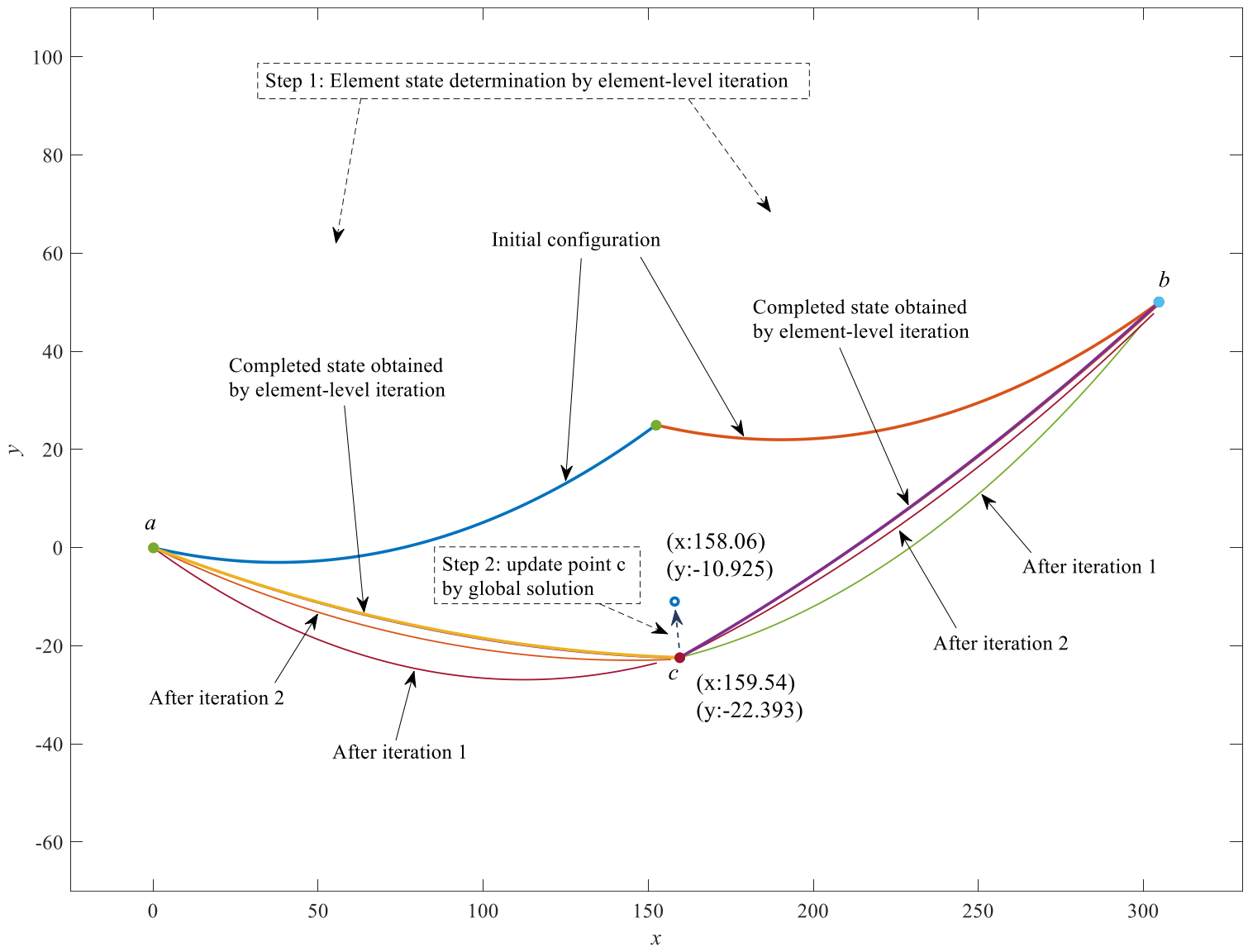

(b) The second update process

**Fig. 8**. Element state determination and nodal position update.

4.1.2 Solution for the problem with given horizontal force

This section validates the applicability of CET-ULS in determining the unstrained length under a specified horizontal force, as depicted in **Fig. 7**b). The specific steps for conducting this validation are outlined below:

Step (a): The completed state obtained by CET-GUL in **Sec. 4.1.1** for each case (refer to **Table 1**, **Table 2** and **Table 3**) is adopted as the initial state for the subsequent solutions. For each case, the horizontal force component at node *b* determined by CET-GUL in **Table 4** is recorded as $H_0$.

Step (b): A numerical model featuring CET-GUL and CET-ULS is constructed and designated as Model-2 (CET). In this model, the left cable element is defined as CET-GUL, while the right one is defined as CET-ULS subjected to the given horizontal force, as illustrated in **Fig. 4**. At the onset of the solution process, the unstrained length of both cable elements is set to half of the distance between points *a* and *b*.

Step (c): With the prescribed horizontal force at the right end having a value of $\lambda H_0$, the solution for the unstrained length of the right element $L_0^{(2)}$ is conducted using Model-2. Subsequently, the total unstrained length of the cable is $L_0 = L_0^{(1)} + L_0^{(2)}$. It should be noted that the unstrained length of the left cable element remains unchanged. In this verification, the coefficient $\lambda$ is set to 0.3 and 1.7, respectively.

Step (d): Reconstruct a numerical model consisting of two CET-GULs, denoted as Model-1 (CET-GUL). Specifically, the unstrained length of both elements is set to $L_0^{(1)} = L_0^{(2)} = L_0/2$. Subsequently, the internal force components at the right end, including *H* and *V*, along with the position of point *c* (*x*, *y*) in the completed state, can be derived by solving Model-1. Based on the validation from the previous section, it can be confirmed that the obtained reaction forces and geometry for a given unstrained length are accurate, thereby providing a reliable standard for comparison in this validation. The relative error of the verification of CET-ULS can then be quantified as

$$\varepsilon = \left\| H - \lambda H_0 \right\| / \left( \lambda H_0 \right) \tag{101}$$

For the CET-ULS, the additional equation can be expressed as

$$h_G = H - \lambda H_0 = 0 \tag{102}$$

Then, the components in Eqs. (67) and (68) are

$$\frac{\partial G}{\partial r_1^a} = \frac{\partial G}{\partial r_2^a} = \frac{\partial G}{\partial r_1^b} = \frac{\partial G}{\partial r_2^b} = \frac{\partial G}{\partial V} = \frac{\partial G}{\partial L_0} = 0,\ \frac{\partial G}{\partial H} = 1 \tag{103}$$

The verification data for the three scenarios of height disparity between two supports are outlined in **Table 5**. The findings indicate that CET-ULS effectively resolves the unstrained length of the cable, affirming the accuracy of the formulation posited in this paper.

**Table 5** Verification of solution for unstrained length.

| Case | $\lambda$ | $\lambda H_0$ | Solved by Model-2 (CET) | | Solved by Model-1 (CET-GUL) | | | $\varepsilon$ |
|---|---|---|---|---|---|---|---|---|
| | | | $L_0^{(2)}$ /m | $L_0$ /m | *H*/kN | *V*/kN | (*x*, *y*) /m | |
| A | 0.3 | 479.99 | 290.085 | 442.485 | 479.99 | 1106.2 | (152.40, -146.88) | 0.00 |
| | 1.7 | 2719.9 | 144.725 | 297.125 | 2719.9 | 742.81 | (152.40, -20.689) | 0.00 |
| B | 0.3 | 553.37 | 252.66 | 407.097 | 553.37 | 1158.0 | (166.72, -94.351) | 0.00 |
| | 1.7 | 3135.8 | 143.842 | 298.278 | 3135.8 | 1268.9 | (155.13, 7.4071) | 0.00 |
| C | 0.3 | 953.94 | 185.962 | 346.354 | 953.94 | 1239.8 | (169.77, -9.8886) | 0.00 |
| | 1.7 | 5405.6 | 137.550 | 297.942 | 5405.6 | 2527.7 | (155.27, 40.456) | 0.00 |

**Fig. 9** exhibits the deformed configuration images for Case B ($l_v = 50\text{m}$), serving to further scrutinize the solution process for the unstrained length of the cable. The figure depicts the configuration state during global iteration, featuring only the configuration changes in element state determination before global iteration for clarity. As depicted in the figure, the unstrained length of the right cable element continuously evolves as the iteration process advances, ultimately converging to a stable outcome. It is noteworthy that the intersection point of the two elements in the figure does not correspond to point *c* after deformation, given that the unstrained lengths of the two elements are no longer equal. Consequently, Model-1 is reconfigured to ascertain the position of point *c*. For the two settings of $\lambda$ for Case B, the total number of global iterations is 8 and 5, respectively. The numbers of element-level iterations for the two elements required in each global iteration, denoted as $N_{\text{ite}}^{(1)}$ and $N_{\text{ite}}^{(2)}$, along with the unstrained length of the right element, are detailed in **Table 6**. Analysis of **Table 6** reveals that the number of iterations needed for element state determination does not exceed 10, and, as the solution progresses, the number of iterations within the element state determination diminishes. These results indicate the favorable convergence of the proposed method in solving implementation.

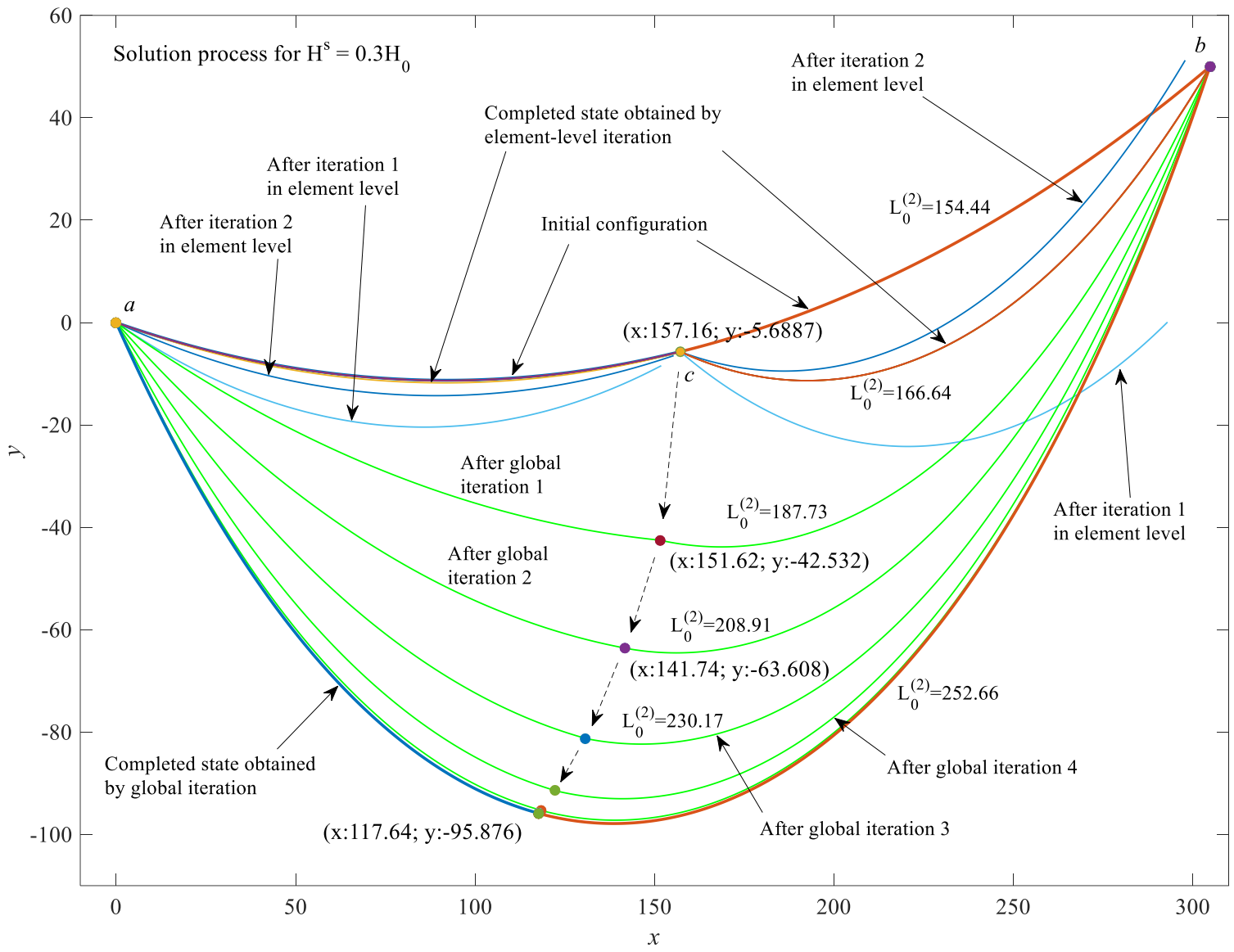

(a) Solution process for $\lambda = 0.3$

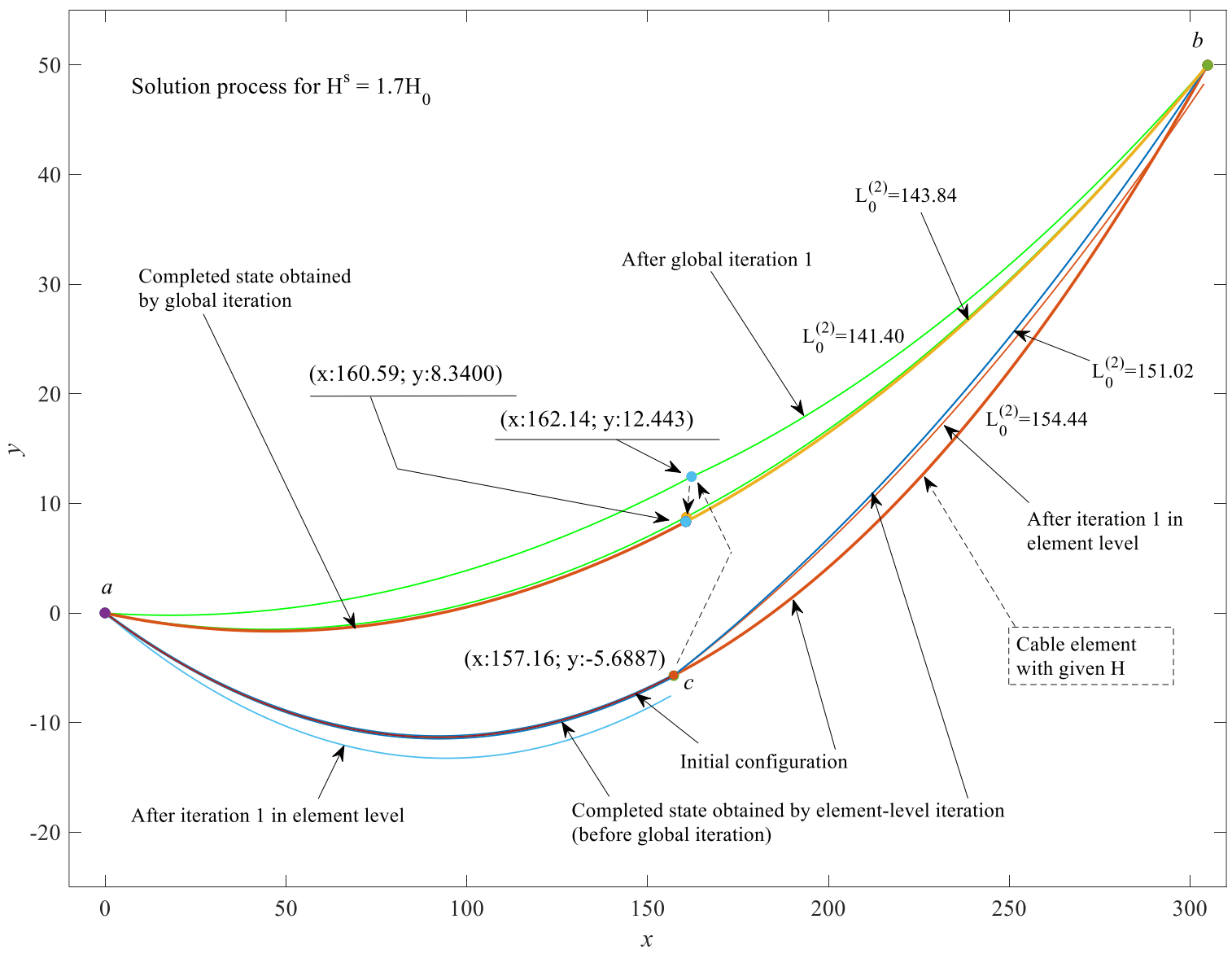

(b) Solution process for $\lambda = 1.7$

**Fig. 9**. Solution process of unstrained length for Case B.

**Table 6** Number of element-level iterations and the unstrained length (Case B).

| Global iteration | $\lambda = 0.3$ | | | $\lambda = 1.7$ | | |
|---|---|---|---|---|---|---|
| | $N_{\text{ite}}^{(1)}$ | $N_{\text{ite}}^{(2)}$ | $L_0^{(2)}$ /m | $N_{\text{ite}}^{(1)}$ | $N_{\text{ite}}^{(2)}$ | $L_0^{(2)}$ /m |
| 1 | 9 | 4 | 187.730 | 6 | 4 | 141.396 |
| 2 | 7 | 4 | 208.905 | 5 | 4 | 143.684 |
| 3 | 6 | 4 | 230.166 | 3 | 3 | 143.841 |
| 4 | 6 | 4 | 245.078 | 2 | 2 | 143.842 |
| 5 | 5 | 3 | 251.579 | 1 | 1 | 143.842 |
| 6 | 4 | 3 | 252.634 | | | |
| 7 | 3 | 2 | 252.660 | | | |
| 8 | 2 | 2 | 252.661 | | | |

Furthermore, analogous verification has been carried out on the cable element subjected to prescribed tension. Given the similarity of the verification process to the work explicated in this section for the element under specified horizontal force, and considering the length constraints of this article, the specific details of the verification for the element under designated tension will not be elaborated. The additional data for the verification can be available on request.

### 4.2 Example 2: Transport pulley system

The examination of the stability of a cable supported by a pulley, previously explored by Bruno and Leonardi [42] and Crusells-Girona et al. [31] is undertaken. **Fig. 10** illustrates the structural model, comprising an inclined cable anchored at both ends and supported by an intermediate roller. The pulley is essentially riding in a frictionless slot and the slot is horizontally set at the location of $y = 100\text{m}$. Then, the location of the pulley could only horizontally change. Meanwhile, the pulley radius is small and hence its influence can be negligible. The cross-sectional area and elastic modulus of the cable are represented as $8.05\times10^{-4}\,\text{m}^2$ and $1.6\times10^{7}\,\text{kN/m}^2$, respectively. Additionally, the unstrained length and the self-weight per unit unstrained length of the cable are specified as $500\text{m}$ and $6.20679\times10^{-2}\,\text{kN/m}$, respectively. This example aims to ascertain the equilibrium configurations of the cable, such that the tension experiences no abrupt change along the roller support, while neglecting friction.

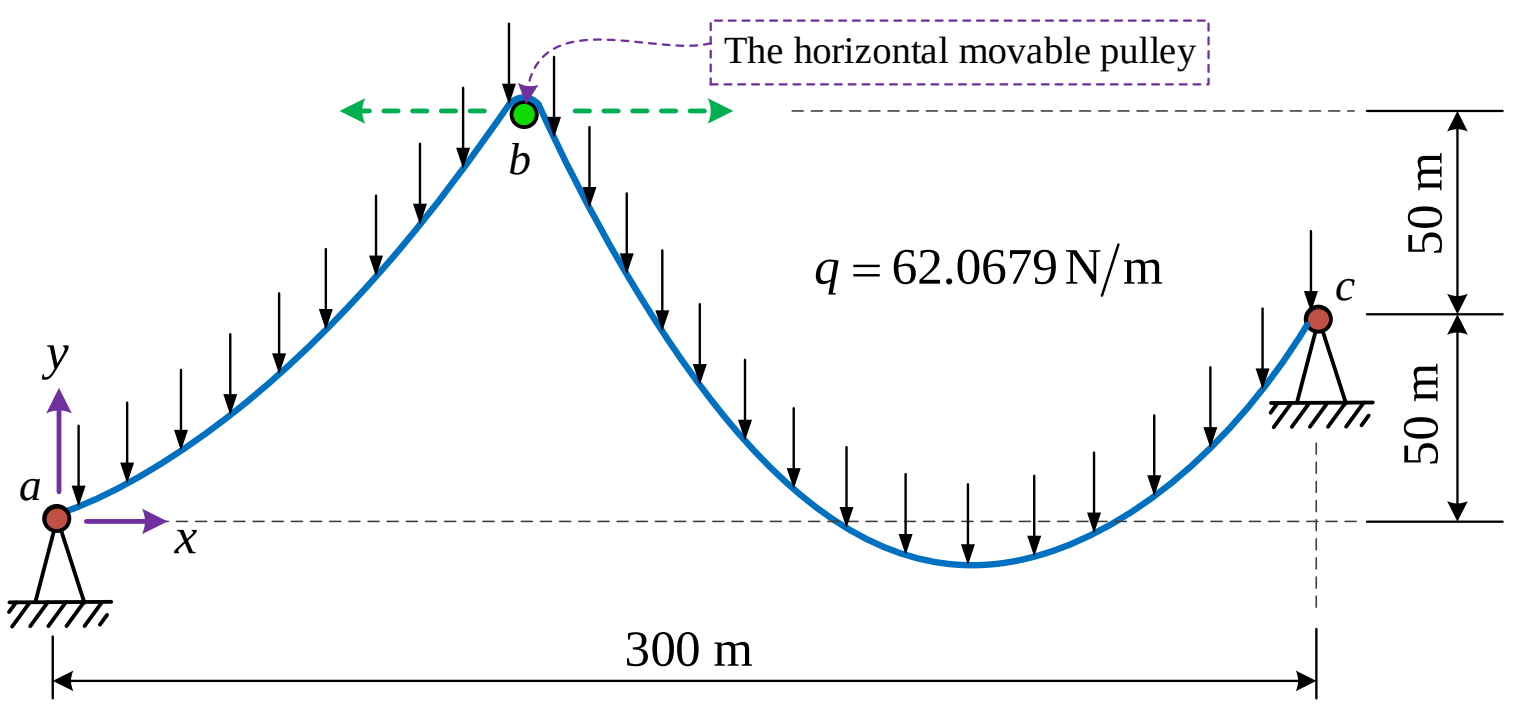

**Fig. 10**. Structural model of the transport pulley system.

The solution to this example encompasses two key aspects: (1) Establishing the correlation between the horizontal position of the pulley and the horizontal reaction at the pulley, assuming continuous cable tension at the pulley location; and (2) Determining the horizontal position of the pulley and the unstrained lengths of the two cable sections, ensuring continuous cable tension and zero horizontal pulley reaction. It is worth noting that the additional constraint equations have to be introduced to describe the relationships of tensions and unstressed lengths between elements. Therefore, the solution method based on complete tangent matrix must be adopted for resolution. To implement the aforementioned solution, a numerical model featuring two CET-ULSs, wherein the unstrained length is considered one of the unknown variables, is constructed. As depicted in **Fig. 11**, the DOFs for element 1 and element 2 are designated as $\left\{ r_1^{(a)} \quad r_2^{(a)} \quad r_1^{(b)} \quad r_2^{(b)} \quad H^{(1)} \quad V^{(1)} \quad L_0^{(1)} \right\}$ and $\left\{ r_1^{(b)} \quad r_2^{(b)} \quad r_1^{(c)} \quad r_2^{(c)} \quad H^{(2)} \quad V^{(2)} \quad L_0^{(2)} \right\}$, respectively. The equations for each cable element are delineated in Eqs. (28)-(30), with the linearized equations presented in Eq. (50). In order to enhance the reader's comprehension of the solution intricacies, a feasible scheme for defining DOFs is illustrated in **Fig. 11**, where the labels in parentheses following each physical quantity denote the DOF sequence of the corresponding quantity in the global system, and 9$^{\text{th}}$-13$^{\text{th}}$ DOFs are constrained DOFs. Specifically, $N_G^s$ represents the tension value at the pulley, considered an unknown of the system, while $F_b$ refers to the horizontal pulley reaction, aligning with the discrepancy in values between the horizontal forces of the two cable elements. Excluding the physical quantities on the constrained degrees of freedom, the system entails a total of 8 unknowns to be solved, namely $r_1^{(b)}$, $H^{(1)}$, $V^{(1)}$, $L_0^{(1)}$, $H^{(2)}$, $V^{(2)}$, $L_0^{(2)}$ and $N_G^s$. However, only 5 equations have been obtained for the system thus far from the equations of the two elements, which pertain to the equilibrium relationship in the horizontal direction at the pulley (1 equation) and the deformation compatibility of the two elements (4 equations). Therefore, three more equations are required. Consequently, three additional equations are necessitated. Considering the relationship between the unstrained lengths of the two elements and the consistency between the tensions of the two elements at the pulley and the provided tension $N_G^s$, the following three equations can be introduced

$$h_g = L_0^{(1)} + L_0^{(2)} - 500 = 0 \tag{104}$$

$$h_G^{(1)} = \left[ H^{(1)} \right]^2 + \left[ V^{(1)} \right]^2 - \left( N_G^s \right)^2 = 0 \tag{105}$$

$$h_G^{(2)} = \left[ H^{(2)} \right]^2 + \left[ V^{(2)} + qL_0^{(2)} \right]^2 - \left( N_G^s \right)^2 = 0 \tag{106}$$

where Eq. (104) guarantees that the total of unstrained lengths of the two elements amounts to 500m, Eqs. (105) and (106) ensure that the tension values of element 1 and element 2 respectively align with the value $N_G^s$ at the pulley. Subsequently, the Taylor series expansion of Eqs. (104)-(106) for the $(i+1)^{\text{th}}$ step in incremental/iterative solution can be articulated as follows

$$h_g^{i+1} \approx h_g^i + \Delta L_0^{(1)} + \Delta L_0^{(2)} = 0 \tag{107}$$

$$h_G^{(1),i+1} \approx h_G^{(1),i} + 2H^{(1)}\Delta H^{(1)} + 2V^{(1)}\Delta V^{(1)} - 2N_G^s \Delta N_G^s = 0 \tag{108}$$

$$h_G^{(2),i+1} \approx h_G^{(2),i} + 2H^{(2)}\Delta H^{(2)} + 2\left(V^{(2)} + qL_0^{(2)}\right)\Delta V^{(2)} + 2q\left(V^{(2)} + qL_0^{(2)}\right)\Delta L_0^{(2)} - 2N_G^s \Delta N_G^s = 0 \tag{109}$$

Subsequently, a comprehensive linearized equation system comprising 8 equations can be formulated to effectuate iterative solution.

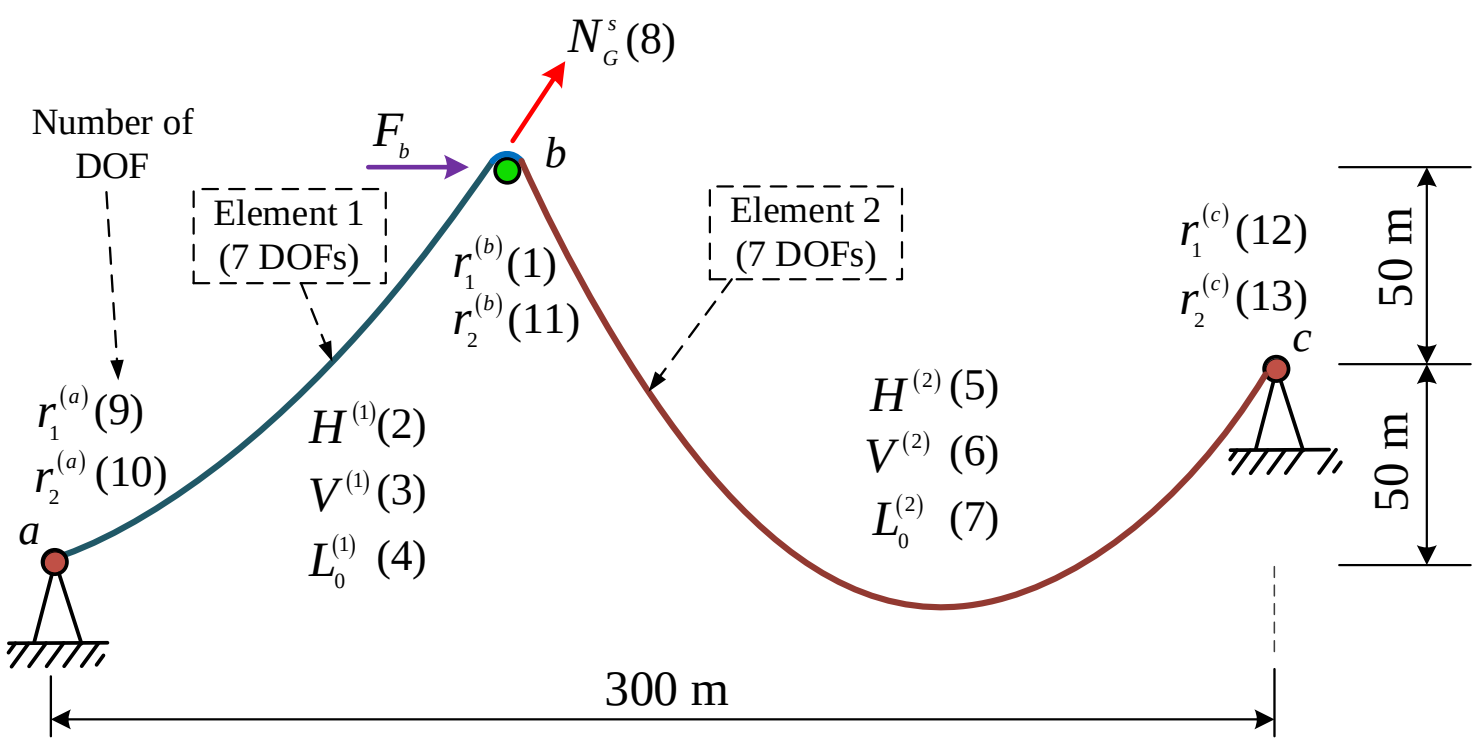

**Fig. 11**. Numerical model of the transport pulley system.

The solution of the relationship between the horizontal position of the pulley and the horizontal pulley reaction proves challenging when employing the iteration method with load control under a given $F_b$, or the iteration method with displacement control under a given horizontal pulley reaction. This difficulty arises from the potential occurrence of multiple states with different $F_b$ that satisfy the continuous cable tension at the pulley within the range of $r_1^{(b)} \in [100.62, 147.00]$, as highlighted in Ref. [32]. Consequently, the incremental/iterative solution method with arc-length control [41] is employed to resolve the relationship between $r_1^{(b)}$ and $F_b$. In this context, $F_b$ is treated as the load factor, and the generalized displacement vector in the arc-length method comprises $r_1^{(b)}$, $H^{(1)}$, $V^{(1)}$, $L_0^{(1)}$, $H^{(2)}$, $V^{(2)}$, $L_0^{(2)}$ and $N_G^s$. **Fig. 12** illustrates the established relationship between $r_1^{(b)}$ and $F_b$, which is consistent with the figure provided by Crusells-Girona et al. [31]. When specifying the arc-length, directly attaining the solution in the equilibrium state with zero horizontal pulley reaction using the arc-length control method proves challenging. Nonetheless, the solution of the arc-length control iteration method facilitates the identification of three points close to the equilibrium state based on the change in sign of the horizontal pulley reaction, as shown by the red points in **Fig. 12**.

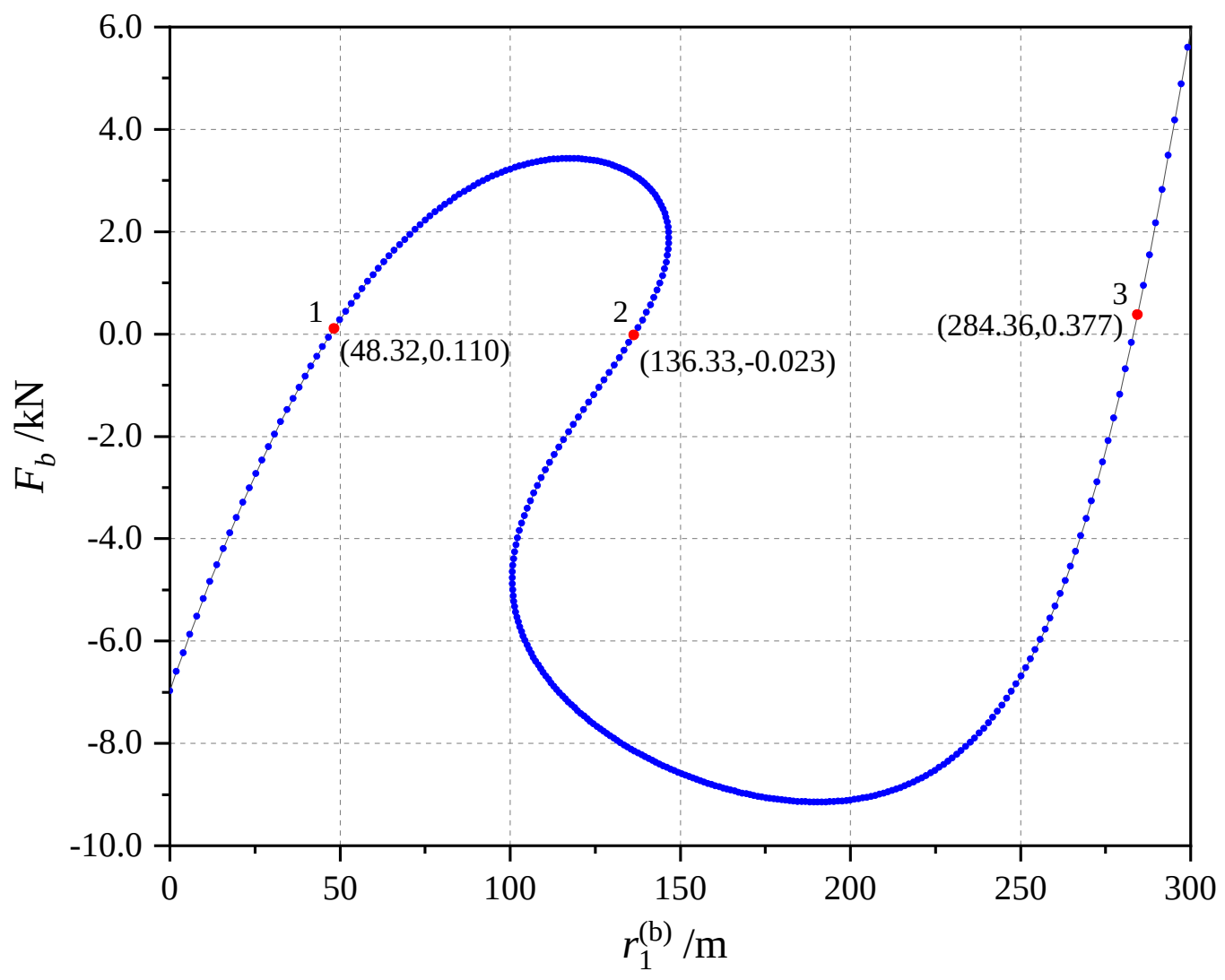

**Fig. 12**. Relation between $r_1^{(b)}$ and $F_b$ obtained by arc-length control method.

Commencing from these three approximate equilibrium points (illustrated as the red points in **Fig. 12**), the equilibrium state with zero horizontal pulley reaction can be ascertained utilizing the iteration method with load control (zero load). The requisite number of iterations for resolving these three equilibrium states does not surpass 3. The unstrained lengths of element 1 $L_0^{(1)}$ and the tension value $N_G^s$ for the three equilibrium states obtained are detailed in **Table 7**, alongside results from other studies. **Table 7** demonstrates that the equilibrium states derived from the method proposed in this paper closely align with those documented in the literature, notably exhibiting high consistency with the findings of Impollonia et al. [32]. These results attest to the capability of the proposed method to address transportation pulley system problems with minimal computational expense, while also validating its efficacy in resolving issues involving undetermined unstrained length.

**Table 7** Results for the equilibrium states from different studies.

| Method | Equilibrium state 1 | | Equilibrium state 2 | | Equilibrium state 3 | |
|---|---|---|---|---|---|---|
| | $L_0^{(1)}$ /m | $N_G^s$ /kN | $L_0^{(1)}$ /m | $N_G^s$ /kN | $L_0^{(1)}$ /m | $N_G^s$ /kN |
| Bruno and Leonardi [42] | 111.07 | 15.499 | - | - | 446.37 | 17.952 |
| Such et al. [28] | 111.96 | 14.531 | - | - | 446.92 | 17.966 |
| Impollonia et al. [32] | 110.83 | 14.531 | 221.52 | 10.631 | 447.30 | 17.982 |
| Crusells-Girona et al. [31] | 110.83 | 14.514 | 221.53 | 10.622 | 447.30 | 17.960 |
| CET-ULS | 110.833 | 14.5309 | 221.518 | 10.6310 | 447.295 | 17.9819 |

To enhance comprehension, **Fig. 13** illustrates the equilibrium configurations and tension distributions for three equilibrium states, with states 1 and 3 representing stable equilibrium, while state 2 denotes an unstable equilibrium, corroborated by observations from **Fig. 12**. As depicted by **Fig. 13**, the horizontal positions of the pulley corresponding to the three equilibrium states are $r_1^{(b)} = 47.254\text{m}$, $r_1^{(b)} = 136.535\text{m}$ and $r_1^{(b)} = 283.149\text{m}$, respectively, and the tension at the pulley is continuous. Utilizing the pulley locations and the determined unstrained lengths corresponding to the three equilibrium states, the numerical model employing adequate truss elements can yield consistent tension outcomes and validate the accuracy of the solution method with CET-ULS.

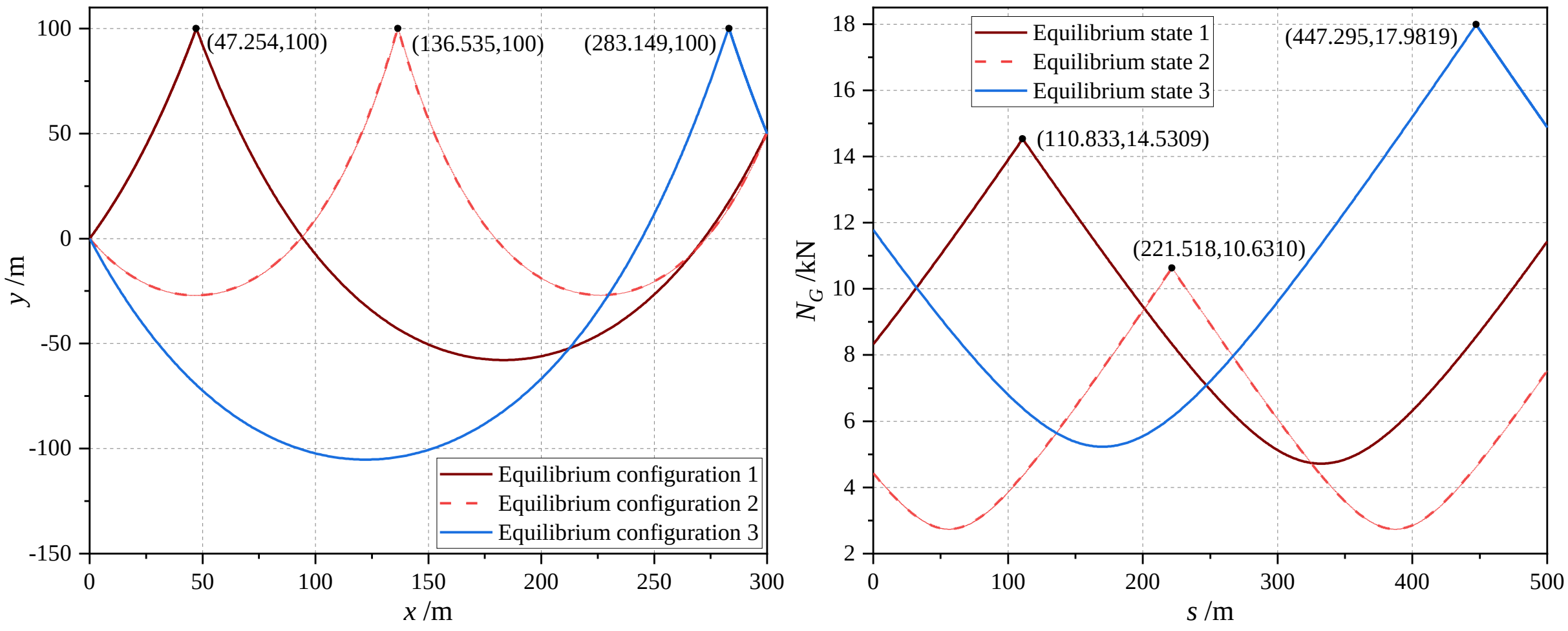

(a) Deformed shape (b) Tension distribution

**Fig. 13**. Deformed shapes of equilibrium states and their tension distributions.

### 4.3 Example 3: Plane cable net

A cable net comprising inclined and horizontal members is analyzed to validate the applicability of CET-GUL in static nonlinear analysis of three-dimensional cable structures. The geometry and initial configuration of the cable structure are illustrated in **Fig. 14**. The cross-sectional area of the cable is 146.45 mm², the elastic modulus is 82,737 MPa, and the self-weight per unit of unstrained length is 1.459 N/m. The unstrained lengths for the inclined and horizontal cables are set at 31.760 m and 30.419 m, respectively. The internal nodes are subjected to a concentrated force of $F_c = 35.586\text{kN}$ . The structure is modeled using 12 cable elements, where each cable is modeled using a single cable element.

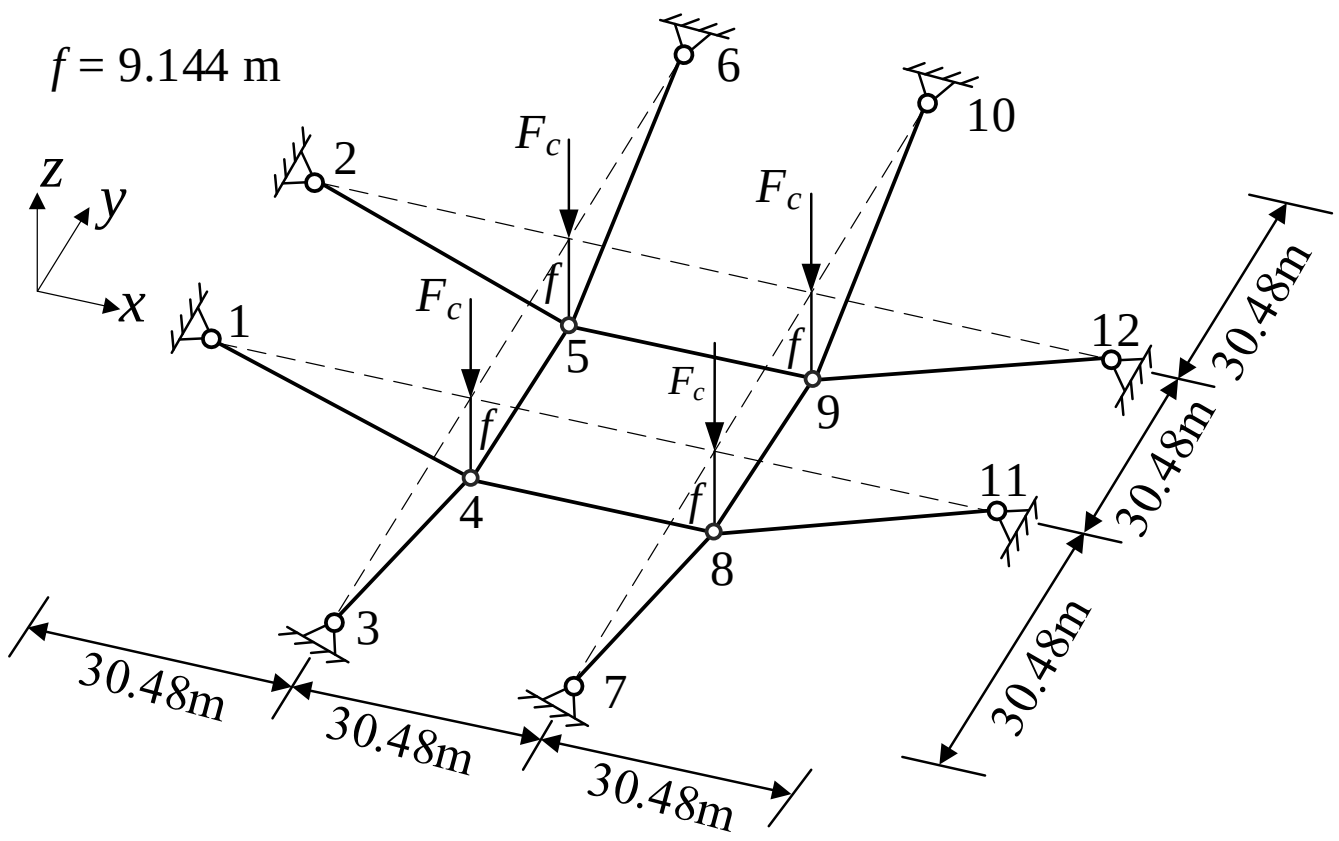

**Fig. 14**. The plane cable net.

The displacements of node 4 (relative to the initial configuration), obtained using CET-GUL and those provided by other researchers, are documented in **Table 8**. As indicated in the table, the results closely align with those from previous studies, suggesting that CET-GUL yields satisfactory outcomes in the static analysis of cable nets.

**Table 8** Comparison of displacements for node 4 in plane cable net.

| Researcher (s) | Method | Displacement at node 4/mm | | |
|---|---|---|---|---|
| | | $x$ -direction | $y$ - direction | $z$ - direction |
| Saafan [43] | Elastic straight | -40.35 | -40.35 | -448.27 |
| Tibert [44] | Elastic catenary | -40.48 | -40.48 | -450.00 |
| Tibert [44] | Associate catenary | -40.78 | -40.78 | -453.36 |
| Thai & Kim [45] | Elastic catenary | -40.13 | -40.13 | -446.50 |
| West & Kar [46] | Nonlinear equilibrium | -40.39 | -40.39 | -447.99 |
| Rezaiee-Pajand et al. [25] | Elastic hyperbolic | -40.75 | -40.75 | -452.79 |
| CET-GUL | Exact tension field | -40.45 | -40.45 | -449.47 |

### 4.4 Example 4: Hyperbolic paraboloid net

A hyperbolic cable network is selected to verify the accuracy of CET-GUL. The selection of this particular case is motivated by the availability of pertinent experimental data, facilitating an objective assessment of the numerical methods employed. As depicted in **Fig. 15**, the structure comprises 31 cable segments that are pretensioned with a force of 200 N before external loads are applied. The internal joints (nodes 5, 6, 7, 10, 11, 12, 15, 16, and 20) are subjected to concentrated loads of 15.7 N. The elastic modulus and the cross-sectional area of all cables are uniformly set at 128.3 GPa and 0.785 mm², respectively. The self-weight per unit unstrained length for all cables is 0.195 N/m. Experimental results for this example are provided by [47].

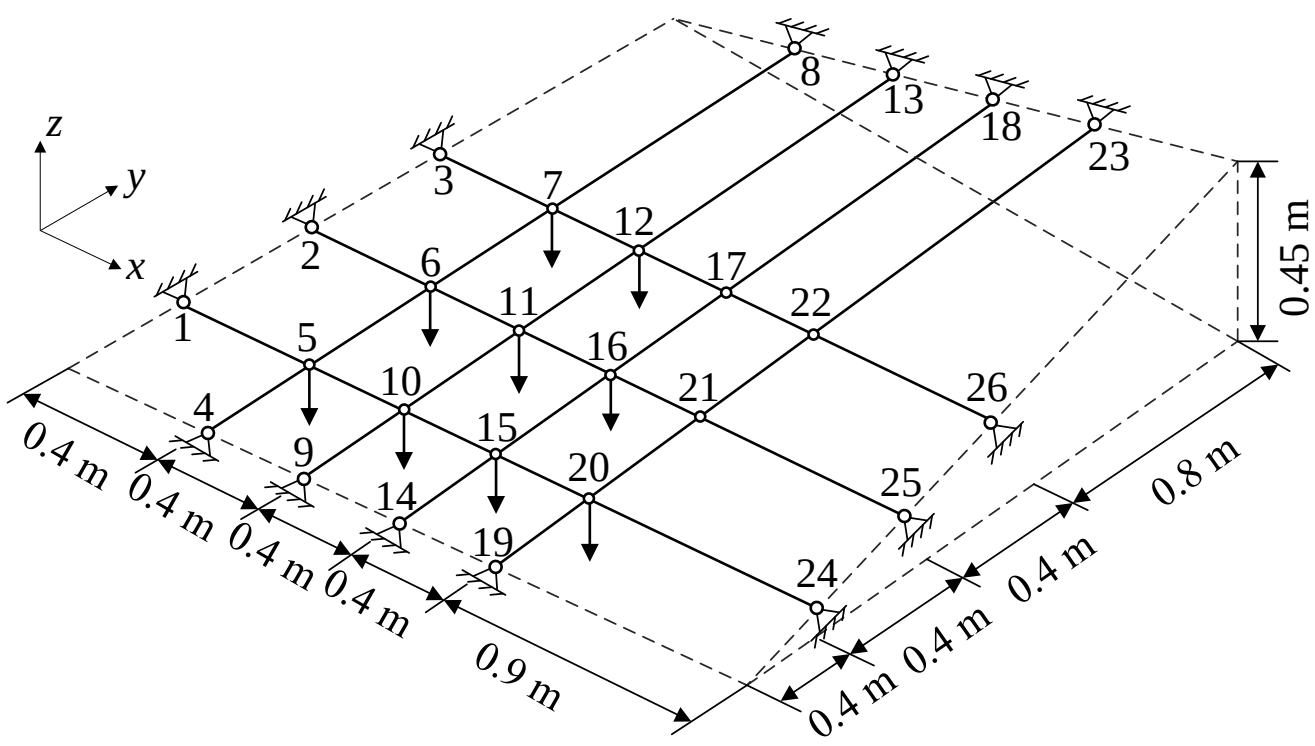

**Fig. 15**. Hyperbolic paraboloid net.

In this study, 31 cable elements are utilized to model the network, with each cable segment represented by a single cable element. Initially, cable elements are defined as straight, and the unstrained length of each cable can be determined as follows prior to load application

$$L_0 = \frac{EA}{P+EA} \times d \tag{110}$$

where $E$ and $A$ represent the elastic modulus and cross-sectional area of the cable, respectively, and $d$ refers to the distance between two nodes of the cable element in the initial configuration.

The vertical displacements of specific nodes, as obtained using CET-GUL, are compared with previous studies in **Table 9**. The values in parentheses within **Table 9** indicate the relative errors (%) compared to experimental results. Notably, the average relative error, defined below, is employed to assess the accuracy of the solution:

$$\text{Error} = \sqrt{\sum_{i \in C} \frac{\left(u_i - u_{i,EXP}\right)^2}{u_{i,EXP}^2}} \times 100\% \tag{111}$$

where $i$ represents the index of node, $C$ refers to the set of internal joints, $u_i$ and $u_{i,EXP}$ represent the vertical displacements of $i$-th node obtained by numerical method and experiment, respectively. As shown in **Table 9**, the displacement solutions derived from CET-GUL closely align with the experimental results. A comparison of average relative errors across various numerical methods reveals that CET-GUL produces the best results among all the listed methods. This demonstrates the CET-GUL's superior accuracy in solution determination.

**Table 9** Comparison of vertical displacements (mm) of hyperbolic paraboloid net.

| Node | Experiment [47] | Dynamic Relaxation [47] | Elastic catenary [45] | Elastic catenary [48] | Elastic hyperbolic [25] | CET-GUL |
|---|---|---|---|---|---|---|
| 5 | -19.50 | -19.30 (1.03) | -19.56 (0.31) | -19.51 (0.05) | -19.51 (0.05) | -19.50 (0.02) |
| 6 | -25.30 | -25.30 (0.00) | -25.70 (1.58) | -25.65 (1.38) | -25.58 (1.10) | -25.56 (1.05) |
| 7 | -22.80 | -23.00 (0.88) | -23.37 (2.50) | -23.37 (2.50) | -23.28 (2.10) | -23.27 (2.04) |
| 10 | -25.40 | -25.90 (1.97) | -25.91 (2.01) | -25.87 (1.85) | -25.83 (1.69) | -25.81 (1.63) |
| 11 | -33.60 | -33.80 (0.60) | -34.16 (1.67) | -34.14 (1.60) | -33.95 (1.04) | -33.94 (1.00) |
| 12 | -28.80 | -29.40 (2.08) | -29.60 (2.78) | -29.65 (2.95) | -29.42 (2.15) | -29.41 (2.10) |
| 15 | -25.20 | -26.40 (4.76) | -25.86 (2.62) | -25.86 (2.62) | -25.61 (1.62) | -25.60 (1.59) |
| 16 | -30.60 | -31.70 (3.59) | -31.43 (2.71) | -31.47 (2.84) | -31.02 (1.37) | -31.00 (1.31) |
| 17 | -21.00 | -21.90 (4.29) | -21.56 (2.67) | -21.57 (2.71) | -21.24 (1.12) | -21.22 (1.07) |
| 20 | -21.00 | -21.90 (4.29) | -21.57 (2.71) | -21.62 (2.95) | -20.83 (0.81) | -20.83 (0.81) |
| 21 | -19.80 | -20.50 (3.54) | -20.14 (1.72) | -20.15 (1.76) | -19.19 (3.08) | -19.18 (3.13) |
| 22 | -14.20 | -14.80 (4.23) | -14.55 (2.46) | -14.55 (2.46) | -13.81 (2.74) | -13.81 (2.74) |
| Error | | 10.63 | 7.81 | 7.94 | 6.14 | 6.06 |

Note: the values in parentheses (.) represent the relative errors (%) with respect to the experiment results.

**4.5 Example 5: Cable with non-uniform cross-sectional stiffness**

This section further validates the solution accuracy of CET-GUL for cables with non-uniform cross-sectional stiffness. The example of single cable as shown in **Fig. 7**a) is used for this validation, and the basic value of cross-sectional stiffness is set to $C_G^0 = 71840.4\text{kN}$. For the cable with non-uniform stiffness, the cross-sectional stiffness along the cable's axis is determined by

$$C_G(s)=\alpha(s)C_G^0 \tag{112}$$

where $\alpha(s)$ represents the stiffness coefficient describing the variation of cross-sectional stiffness along the axis of the cable. For $\alpha(s)$, two scenarios including linear distribution (I) and quadratic distribution (II) are defined as

$$\alpha(s)=\begin{cases}\alpha_1+\dfrac{s}{L_0}(\alpha_2-\alpha_1) & \text{for linear distribution (I)}\\ \dfrac{4(\alpha_2-\alpha_1)}{L_0^2}s^2+\dfrac{4(\alpha_1-\alpha_2)}{L_0}s+\alpha_2 & \text{for quadratic distribution (II)}\end{cases} \tag{113}$$

where $L_0$ represents the unstrained length of the cable, and $\alpha_1$ and $\alpha_2$ refer to the given parameters determining the distribution of $\alpha(s)$.

Assuming an unstressed cable length of 308.8 m and setting parameters as $\alpha_1=0.5$ and $\alpha_2=1.5$, the coordinates (*x*, *y*) of point *c* (refer to **Fig. 7**a)) in the equilibrium states under two scenarios are obtained by $\alpha(s)$ using both CET-GUL and the truss element (TRUSS). These coordinates are detailed in **Table 10**, **Table 11** and **Table 12**, respectively, for three different cases of height differences between the two supports (as defined in Eq. (100)). Particularly, results under various refinements are provided to illustrate their convergence trend. The findings confirm that CET-GUL is highly suitable for cables with non-uniform cross-sectional stiffness and demonstrates exceptional solution accuracy, thus validating the formulation presented in this paper.

**Table 10** Location (*x*, *y*) of point *c* obtained by the four numerical models (Case A).

| $N_e$ | (*x*, *y*) /m | | | |
|---|---|---|---|---|
| | Linear distribution (I: $\alpha_1=0.5,\alpha_2=1.5$) | | Quadratic distribution (II: $\alpha_1=0.5,\alpha_2=1.5$) | |
| | TRUSS | CET-GUL | TRUSS | CET-GUL |
| 1 | - | 153.44, -37.040 | - | 152.40, -38.837 |
| 2 | 153.21, -39.907 | 153.44, -37.040 | 152.40, -41.783 | 152.40, -38.837 |
| 4 | 153.39, -37.738 | | 152.40, -39.604 | |
| 8 | 153.43, -37.231 | | 152.40, -39.079 | |
| 16 | 153.44, -37.089 | | 152.40, -38.905 | |
| 32 | 153.44, -37.052 | | 152.40, -38.854 | |
| 64 | 153.44, -37.043 | | 152.40, -38.841 | |
| 128 | 153.44, -37.041 | | 152.40, -38.838 | |
| 256 | 153.44, -37.040 | | 152.40, -38.838 | |
| 512 | 153.44, -37.040 | | 152.40, -38.837 | |

**Table 11** Location (*x*, *y*) of point *c* obtained by the four numerical models (Case B).

| $N_e$ | (*x*, *y*) /m | | | |
|---|---|---|---|---|
| | Linear distribution (I: $\alpha_1 = 0.5, \alpha_2 = 1.5$) | | Quadratic distribution (II: $\alpha_1 = 0.5, \alpha_2 = 1.5$) | |
| | TRUSS | CET-GUL | TRUSS | CET-GUL |
| 1 | - | 158.45, -6.4077 | - | 157.62, -8.8277 |
| 2 | 158.59, -7.8850 | 158.45, -6.4077 | 157.95, -10.015 | 157.62, -8.8277 |
| 4 | 158.48, -6.7777 | | 157.69, -9.0917 | |
| 8 | 158.46, -6.5262 | | 157.64, -8.9583 | |
| 16 | 158.46, -6.4427 | | 157.63, -8.8807 | |
| 32 | 158.45, -6.4165 | | 157.62, -8.8411 | |
| 64 | 158.45, -6.4098 | | 157.62, -8.8309 | |
| 128 | 158.45, -6.4081 | | 157.62, -8.8285 | |
| 256 | 158.45, -6.4078 | | 157.62, -8.8279 | |
| 512 | 158.45, -6.4077 | | 157.62, -8.8277 | |

**Table 12** Location (*x*, *y*) of point *c* obtained by the four numerical models (Case C).

| $N_e$ | (*x*, *y*) /m | | | |
|---|---|---|---|---|
| | Linear distribution (I: $\alpha_1 = 0.5, \alpha_2 = 1.5$) | | Quadratic distribution (II: $\alpha_1 = 0.5, \alpha_2 = 1.5$) | |
| | TRUSS | CET-GUL | TRUSS | CET-GUL |
| 1 | - | 159.75, 32.838 | - | 158.74, 29.467 |
| 2 | 159.41, 33.432 | 159.75, 32.838 | 158.07, 31.609 | 158.74, 29.467 |
| 4 | 159.56, 33.266 | | 158.31, 30.861 | |
| 8 | 159.67, 33.026 | | 158.57, 30.038 | |
| 16 | 159.74, 32.870 | | 158.72, 29.547 | |
| 32 | 159.75, 32.838 | | 158.75, 29.464 | |
| 64 | 159.75, 32.838 | | 158.74, 29.465 | |
| 128 | 159.75, 32.838 | | 158.74, 29.466 | |
| 256 | 159.75, 32.838 | | 158.74, 29.467 | |
| 512 | 159.75, 32.838 | | 158.74, 29.467 | |

## 5 Conclusions

This paper presents a numerically exact cable finite element model for static nonlinear analysis of cable structures. Within this model, the tension field is precisely derived using geometrically exact beam theory alongside the fundamental mechanical properties of cables. The model formulates the cable finite element considering given and to-be-solved unstrained lengths, by deriving linearized equations with implicit integral expressions. In addition, two solution methods are implemented. The method with element internal iterations ensures the regularity of the numerical model, and the method with complete tangent matrix is applied to cables with additional constraint equations, which describe the relationships of force parameters and unstrained lengths between elements. Numerical examples validate the effectiveness and computational efficiency of the proposed cable finite elements and its solution methods. The following conclusions are drawn:

(1) The proposed cable finite elements, based on an exact representation of the tension field, demonstrate exceptional accuracy, typically requiring only one element per cable segment for precise results.

(2) The accuracy of the finite element formulation, considering both specified and to-be-solved

unstrained lengths, supports efficient implementation of iterative solutions in nonlinear analysis of cable structures.

(3) The proposed finite element effectively tackles the challenge of determining the state of a cable with an unsolved unstrained length, highlighting the broad applicability of the element.

(4) The proposed elements are capable of modeling the static nonlinear behavior of cables featuring non-uniform cross-sectional stiffness. This capability enables potential applications in engineering computations involving varying cross-sectional stiffness, such as uncertainty analysis and structural optimization.

(5) By incorporating an iterative algorithm with arc-length control and additional control conditions, the cable elements can address complex problems, such as analyzing the relationship between pulley position and horizontal pulley reaction in a transport pulley system.

While the proposed cable elements are effective for static nonlinear analysis of cable structures, two limitations require further consideration. Firstly, these elements cannot accommodate distributed horizontal loads, posing challenges in accurately simulating the nonlinear behavior of cable structures under wind loads. Secondly, despite their high performance in static analysis, these elements are not applicable to dynamic behavior analysis of cable structures, an area that demands in-depth investigation in engineering. Consequently, future research should focus on developing high-performance cable elements that can handle distributed horizontal loads and are suitable for dynamic analysis.

## Acknowledgments

The project is funded by the National Natural Science Foundation of China (Grant No. 52178209, Grant No. 51878299) and Guangdong Basic and Applied Basic Research Foundation, China (Grant No. 2021A1515012280, Grant No. 2020A1515010611).